\documentclass[aps,pra,reprint,amsmath,amssymb,longbibliography,floatfix]{revtex4-2}
\usepackage[T1]{fontenc}
\usepackage[english]{babel}
\usepackage{lmodern}
\usepackage{bm,graphicx,booktabs,xcolor}
\usepackage{needspace}
\usepackage{placeins}
\usepackage{array}[=2016-10-06]
\usepackage[colorlinks=true,linkcolor=black,citecolor=black,urlcolor=black]{hyperref}

\hypersetup{hidelinks}

\makeatletter
\let\arxivNormalHangFromSection\@hangfrom@section
\let\arxivNormalSectionCountFormat\@sectioncntformat
\makeatother
\begin{document}
\title{Unbiased Estimation of Conditional Covariance for Quantum Optomechanics}
\author{Katsuta Sakai}
\affiliation{Department of Physics, Faculty of Science, Gakushuin University, 1-5-1 Mejiro, Toshima, Tokyo 171-8588, Japan}
\affiliation{TRIP Headquarters, RIKEN, Wako 351-0198, Japan}
\author{Nobuyuki Matsumoto}
\email{nobuyuki.matsumoto@gakushuin.ac.jp}
\affiliation{Department of Physics, Faculty of Science, Gakushuin University, 1-5-1 Mejiro, Toshima, Tokyo 171-8588, Japan}
\begin{abstract}
Continuous measurements can prepare macroscopic mechanical oscillators in conditional quantum states, but their covariance is difficult to verify. For each quadrature, the conventional retrodictive estimator is unbiased exactly when the conditional and future-likelihood variances are equal, and it needs a separate rule for the position--momentum cross covariance. Here, we derive an estimator that recovers all three independent covariance entries using one exact identity for linear-Gaussian models. It combines the forward trajectory, a future-likelihood center, and the smoothed trajectory. Using these three trajectories, we reconstruct the covariance of the center-of-mass motion of a $7.71$-mg suspended mirror. This covariance describes a thermally squeezed conditional state with a principal-variance ratio of $4.7\pm0.7$. The reconstructed covariance and the measured bias difference between the conventional and three-trajectory estimators agree with the model predictions within propagated uncertainties. Varying the upper band edge resolves changes in all three covariance entries. For upper edges up to $1\,\mathrm{kHz}$, the measured changes agree with the finite-band predictions. Every experimental record covers only a finite frequency band; these results show how such records can be used to characterize conditional mechanical states and provide a basis for future tests of macroscopic entanglement.
\end{abstract}
\maketitle
\section{Introduction}

Continuous monitoring turns quantum measurement backaction into a resource for preparing and controlling mechanical motion. Such a measurement conditions the mechanical state on the detector record, producing a record-dependent mean and residual conditional fluctuations. The record can then be processed offline to reconstruct this conditional motion or in real time to drive feedback and modify the unconditional dynamics~\cite{Wieczorek2015,Wilson2015,Rossi2018,Rossi2019}. Conditioning on the record is particularly attractive for macroscopic oscillators for two reasons. First, following each measurement-conditioned trajectory removes the spread of the trajectory means and leaves only the residual conditional uncertainty~\cite{Rossi2019}. Second, it provides a viable pathway to macroscopic quantum entanglement between two mirrors, which has been proposed for platforms ranging from milligram- to kilogram-scale pendulums~\cite{MuellerEbhardt2008,Miki2023}.

Indeed, within a calibrated dynamical model, this conditioning can reveal a localized, nearly pure state even when the unconditional ensemble is strongly mixed by thermal environments. Optomechanical experiments have observed conditional mechanical trajectories~\cite{Rossi2019}, conditional mechanical squeezing~\cite{Meng2022}, and sub-phonon occupation of a collective mechanical mode~\cite{Huang2024}. Theoretically, in the linear-Gaussian regime, the conditional mean obeys a quantum Kalman--Bucy equation and the conditional covariance follows a deterministic Riccati equation~\cite{Wieczorek2015,KalmanBucy1961}. For Gaussian states, the full covariance matrix in turn determines the purity and, for a bipartite Gaussian state, its entanglement properties.

A central challenge is therefore verifying the conditional covariance using an experimental record. The forward trajectory, inferred from measurements up to a given time, specifies only the conditional mean. Assessing its uncertainty also requires separating mechanical fluctuations from the noise in the detector output. For this purpose, a two-trajectory approach constructs a second estimate from subsequent measurements, motivated by past-quantum-state retrodiction~\cite{Gammelmark2013,ZhangMolmer2017} and implemented experimentally by Rossi \textit{et al.}~\cite{Rossi2019}. Here, the likelihood of the subsequent record, considered as a function of the state at the current time, is the future likelihood. Its center gives the second estimate, called the future-only likelihood center.

The variance of the difference between this center and the forward trajectory contains both the conditional covariance and the width of this likelihood. The conventional prescription halves this difference variance. However, for the future-only likelihood used here, this gives the conditional variance only when the corresponding conditional and future-likelihood variances are equal. The limitations of this prescription and the associated reconstruction bias have been studied in subsequent work~\cite{Meng2022,Hatakeyama2026}. A related suspended-mirror analysis is discussed in Refs.~\cite{SantiagoCondori2020,SantiagoCondori2020_note}. 

The position--momentum cross covariance poses an additional challenge: it determines the orientation of conditional squeezing and enters Gaussian entanglement criteria~\cite{Miki2023}. Meng \textit{et al.}~\cite{Meng2022} and Hatakeyama \textit{et al.}~\cite{Hatakeyama2026} analyzed oscillator models using causal and anticausal Wiener estimates, based on the past and future records, respectively. In those models, the difference between the two estimates has zero position--momentum covariance, although the conditional cross covariance is generally nonzero. Those studies therefore used a separate prescription for the conditional cross covariance.

To overcome these limitations, we introduce a third, smoothed trajectory that uses both past and future measurements~\cite{Rauch1965,Mayne1966}. For linear-Gaussian models, the resulting three-trajectory reconstruction recovers all three independent covariance entries from trajectory-difference statistics through an identity that allows the conditional covariance and future-likelihood width to differ. Although the same relation is used in data assimilation~\cite{Desroziers2005,Todling2022} (Appendix~\ref{sec:identity_proof}), this work is, to our knowledge, its first application to an optomechanical experiment. Beyond this application, we derive the bias of the conventional prescription under colored force noise. For a variant that conditions the future likelihood on the past record, we also derive an analytic bound on the reconstruction error from frequencies outside the analysis band.

We then apply the method to the detector record from a suspended-mirror experiment~\cite{Matsumoto2019}, using the measured mirror mass of $7.71\,\mathrm{mg}$~\cite{SantiagoCondori2020} (SM Table~\ref{tab:sm_upstream}) to reconstruct its center-of-mass motion. The reconstruction characterizes a thermally squeezed conditional state. In this analysis, we use a model built from measured device parameters, noise characteristics, and transfer functions. The reconstructed covariance agrees with the model predictions within the propagated uncertainties. So does the measured bias difference, the difference between the outputs of the two prescriptions applied to the same record. Because the analysis uses a finite frequency band, the model also predicts how the reconstruction depends on that band. We therefore vary the analysis band and compare the measured covariance changes with the finite-band predictions. This record is dominated by classical noise (thermal force and laser frequency noise), so the experiment tests the method on a conditional state far above the vacuum level. Nevertheless, the method applies equally when quantum noise dominates.

Section~\ref{sec:theory} derives the identity, the conventional bias, and the bandwidth criterion. Section~\ref{sec:experiment} describes the apparatus, calibration, and processing, and Sec.~\ref{sec:results} presents the comparisons. Sections~\ref{sec:discussion} and \ref{sec:conclusions} discuss the scope and summarize the results. Appendixes~\ref{app:identity}--\ref{app:bandwidth} give the proofs and ideal-model examples; the Supplemental Material (SM)~\cite{SupplementalMaterial} documents the physical model and the uncertainty calculation.

\section{Covariance reconstruction from three trajectories}
\label{sec:theory}

We first define the conditional covariance and the three trajectories used to reconstruct it. We then derive the reconstruction identity, evaluate the conventional bias, and bound the effect of a finite analysis band.
The list below summarizes the main symbols.

\par\medskip
\begingroup
\noindent\textbf{Main symbols}\par\nobreak
\small\raggedright
\newcommand{\symbolgroup}[1]{\par\smallskip\noindent\textit{#1}\par\nobreak}
\newcommand{\symbolentry}[2]{%
  \par\noindent\hangindent=0.24\columnwidth\hangafter=1
  \makebox[0.24\columnwidth][l]{#1}#2\par}
\symbolgroup{Trajectories of the mechanical coordinates}
\symbolentry{$\bm f$}{Forward trajectory from the past record [Eq.~\eqref{eq:theory_target}].}
\symbolentry{$\bm b_0$}{Future-only likelihood center from the future record [Eq.~\eqref{eq:original_center}].}
\symbolentry{$\bm s$}{Smoothed trajectory: the conditional mean given both records [Eq.~\eqref{eq:augmented_smoother}].}
\symbolentry{$\bm b_c$}{Center of the mechanical future likelihood conditioned on the past record [Eq.~\eqref{eq:main_conditional_center}].}
\symbolentry{$\bm a$}{Trajectory difference $\bm f-\bm s$.}
\symbolgroup{Covariances}
\symbolentry{$V$}{Conditional covariance given the past record; the reconstruction target [Eq.~\eqref{eq:theory_target}].}
\symbolentry{$V^{(\ell)}$}{Width of the future likelihood [Eq.~\eqref{eq:original_center}].}
\symbolentry{$U$}{Smoothing-error covariance [Eq.~\eqref{eq:augmented_smoother}].}
\symbolentry{$D$}{$V-U=\operatorname{Cov}(\bm a)=C_{sf}$ [Eqs.~\eqref{eq:main_conditional_center}--\eqref{eq:difference_covariance_values}].}
\symbolentry{$C_{fb},C_{sf},C_{sb}$}{Covariances of trajectory differences [Eq.~\eqref{eq:difference_covariances}].}
\symbolentry{$P$}{Augmented-state forward error covariance [Eq.~\eqref{eq:theory_riccati}].}
\symbolgroup{Predictions and experimental statistics}
\symbolentry{$V_2$}{Full-band prediction of the conventional prescription [Eq.~\eqref{eq:conventional_rule}].}
\symbolentry{$P_{3,B},P_{3c,B}$}{Finite-band three-trajectory predictions with $\bm b_0$ and $\bm b_c$ [Eqs.~\eqref{eq:theory_band_integral} and \eqref{eq:conditional_band_prediction}].}
\symbolentry{$P_2,P_3,P_{3c}$}{Finite-band predictions for the experimental band, with $P_3=P_{3,B}$ and $P_{3c}=P_{3c,B}$ (Sec.~\ref{sec:experiment}\,C).}
\symbolentry{$E_2,E_3,E_{3c}$}{Corresponding experimental statistics (Sec.~\ref{sec:experiment}\,C). A tilde on $E$ means that we have subtracted the mean offset due to finite-record processing, estimated from simulated records (Sec.~\ref{sec:results}\,A).}
\symbolgroup{Noise intensities}
\symbolentry{$\mathcal Q,\mathcal R,\mathcal L$}{State-noise, readout-noise, and cross-covariance intensities [Eq.~\eqref{eq:theory_noise}].}
\symbolentry{$R_y$}{Value of $\mathcal R$ for the readout $dY=q\,dt+dw$ (Sec.~\ref{sec:colored_bias_mechanism}).}
\symbolgroup{Analysis band}
\symbolentry{$B$}{Analysis band, as in $P_{3,B}$ and $P_{3c,B}$.}
\symbolentry{$\epsilon_B,\epsilon_{B,c}$}{Relative band-truncation errors with $\bm b_0$ and $\bm b_c$ [Eq.~\eqref{eq:band_accuracy}].}
\symbolentry{$\zeta_B,g_c$}{Largest omitted fraction of $D$ and an upper bound on its amplification [Eq.~\eqref{eq:band_capture_metric}].}
\endgroup

\subsection{The conditional mechanical covariance}

We consider the uncertainty in the mechanical state remaining after the detector record up to the current time has been used. For the linear-Gaussian model, we use continuous-time Kalman--Bucy filtering for the past record and fixed-interval smoothing for the combined past and future record~\cite{Bensoussan1992,SarkkaSvensson2023}. The model includes a force noise that is correlated in time (colored force noise) and noise inputs that both drive the motion and enter the readout. Let $\bm x=(q,p)^{\mathsf T}$ denote the dimensionless mechanical quadratures, normalized so that $[q,p]=\mathrm i$ and the vacuum covariance is $I_2/2$, where $I_2$ is the two-dimensional identity matrix.

\begingroup\displaywidowpenalty=10000
For the suspended mirror considered here, structural damping describes internal material losses in the suspension and provides the reference model for colored thermal-force noise~\cite{Saulson1990,GonzalezSaulson1994}. In the classical high-temperature regime, a nearly frequency-independent loss angle, which describes the dissipative part of the elastic response, gives a thermal-force power spectral density (PSD) approximately proportional to $1/\nu$ at positive frequency $\nu$ over its applicable range. Throughout, power and cross-spectral densities are one-sided densities per hertz.

To represent the temporal correlations of the colored force, we approximate the colored-force spectrum over the modeled frequency range by auxiliary Ornstein--Uhlenbeck (OU) processes~\cite{UhlenbeckOrnstein1930}, collected in $\bm\eta$. Each OU process is driven by white noise and has an exponentially decaying temporal correlation. The mechanical coordinates together with these variables therefore form an augmented Markov state.

The augmented state $\bm z=(\bm x^{\mathsf T},\bm\eta^{\mathsf T})^{\mathsf T}$ obeys the linear-Gaussian model
\begin{equation}
 d\bm z=A\bm z\,dt+d\bm v,
 \qquad dY=C\bm z\,dt+dw .
 \label{eq:theory_state}
\end{equation}
Here $A$ is the drift matrix, $C$ is the measurement matrix, $Y$ is the accumulated detector record, and $d\bm v,dw$ are zero-mean Gaussian increments.

Their covariance matrices are
\begin{equation}
 \mathbb E\!\left[
 \begin{pmatrix}d\bm v\\dw\end{pmatrix}
 \begin{pmatrix}d\bm v\\dw\end{pmatrix}^{\mathsf T}
 \right]
 =\begin{pmatrix}\mathcal Q&\mathcal L\\
 \mathcal L^{\mathsf T}&\mathcal R\end{pmatrix}dt .
 \label{eq:theory_noise}
\end{equation}
For operator-valued inputs, each expectation in Eq.~\eqref{eq:theory_noise} denotes the symmetrically ordered product; for classical inputs it is the usual product. The matrix $\mathcal Q$ specifies diffusion of the state, $\mathcal R$ specifies the detector noise, and $\mathcal L$ is the cross covariance between the state noise and the detector noise. For example, an optical intensity fluctuation can both exert radiation pressure and enter the detected intensity. Using the same input in these two paths produces $\mathcal L$ and determines its sign. For the single output considered here, $\mathcal R$ is a scalar and $\mathcal L$ is a column vector with one entry per state variable.\par\endgroup

At time $t$, let $\mathcal Y_-$ contain the record through $t$ and let $\mathcal Y_+$ contain the subsequent record through its endpoint $T$. The forward mean $\bm f_z=\mathbb E[\bm z_t\mid\mathcal Y_-]$ and its augmented-state error covariance $P$ follow the Gaussian filtering equations~\cite{Wieczorek2015,KalmanBucy1961,Bensoussan1992},
\begin{align}
 d\bm f_z&=A\bm f_z\,dt+K(dY-C\bm f_z\,dt),\notag\\
 K&=(PC^{\mathsf T}+\mathcal L)\mathcal R^{-1},\notag\\
 \dot P&=AP+PA^{\mathsf T}+\mathcal Q-K\mathcal R K^{\mathsf T}.
 \label{eq:theory_riccati}
\end{align}
Here $K$ is the Kalman gain, which weights the innovation $dY-C\bm f_z\,dt$: the measured signal in excess of that predicted from the past.

With $\Pi=(I_2\;0)$ selecting the mechanical coordinates, the forward mechanical trajectory and conditional covariance are
\begin{equation}
 \bm f=\Pi\bm f_z,\qquad V=\Pi P\Pi^{\mathsf T}.
 \label{eq:theory_target}
\end{equation}
Here $V$ is the covariance of the mechanical state conditioned on the past record. Let $\rho_c(t)$ be that state. Let $\hat q$ and $\hat p$ denote the mechanical quadrature operators. The trajectory components are $f_q=\operatorname{Tr}[\rho_c(t)\hat q]$ and $f_p=\operatorname{Tr}[\rho_c(t)\hat p]$. With $\delta\hat q=\hat q-f_q$ and $\delta\hat p=\hat p-f_p$, the covariance is

\begin{equation}
 V=\operatorname{Tr}\!\left[\rho_c(t)
 \begin{pmatrix}
 (\delta\hat q)^2 & \tfrac12\{\delta\hat q,\delta\hat p\}\\
 \tfrac12\{\delta\hat q,\delta\hat p\} & (\delta\hat p)^2
 \end{pmatrix}\right],
 \label{eq:quantum_conditional_covariance}
\end{equation}
where $\{\hat O_1,\hat O_2\}=\hat O_1\hat O_2+\hat O_2\hat O_1$ defines the anticommutator of two operators. Thus $V_{qq}$ and $V_{pp}$ are the conditional position and momentum variances, and $V_{qp}=V_{pq}$ is their conditional cross covariance.
Symmetric ordering gives a real phase-space covariance. For a linear quantum system, these moments obey the same filtering equations~\eqref{eq:theory_riccati} as for a classical linear system, so Eqs.~\eqref{eq:theory_target} and \eqref{eq:quantum_conditional_covariance} give the same $V$, while the physical input fields constrain the allowed noise matrices~\cite{Wieczorek2015}. For stationary comparisons, we obtain $V$ by projecting the stationary Riccati solution $P$ onto the mechanical coordinates.

To illustrate the role of the cross covariance, consider the pure Gaussian covariance in the present normalization
\begin{equation}
 V_{\rm sq}=\frac18\begin{pmatrix}5&-3\\-3&5\end{pmatrix}.
 \label{eq:squeezed_covariance_example}
\end{equation}
Its eigenvalues are $1/4$ and $1$, and its determinant is $1/4$, as required for a pure one-mode Gaussian state. The rotated quadrature $(q+p)/\sqrt2$ has variance $1/4$, or approximately $3$ dB below vacuum, even though both diagonal entries are $5/8>1/2$. Setting the cross covariance to zero would erase this squeezing. Accordingly, all filtering, uncertainty propagation, and matrix comparisons below include the cross-covariance entries.

\subsection{Future information in the augmented state}
\label{sec:future_information}

We combine the forward trajectory $\bm f$ and the smoothed trajectory $\bm s$, which uses both records, with either of two future-likelihood centers. The future-only likelihood center $\bm b_0$ uses the future record alone. The past-conditioned center $\bm b_c$ uses the future record together with what the past record tells us about the auxiliary force variables. We use $\bm b_0$ to compare the two prescriptions, because the conventional prescription is defined with it, and $\bm b_c$ to compare the reconstruction with $V$, because its finite-band error is bounded (Sec.~\ref{sec:ideal_bandwidth}).

\textit{Future-only likelihood center.} Offline processing makes the record after $t$, $\mathcal Y_+$, available. What it implies about the state at $t$ is described by a Gaussian future likelihood~\cite{Mayne1966},
\begin{equation}
 p(\mathcal Y_+\mid\bm z_t)\propto
 \exp[-\tfrac12\bm z_t^{\mathsf T}J\bm z_t+
 \bm h^{\mathsf T}\bm z_t].
 \label{eq:augmented_future}
\end{equation}
Here $J$ and $\bm h$ are the information matrix and vector: the quadratic and linear coefficients of the Gaussian log likelihood~\cite{SarkkaSvensson2023}. Both vanish at the record endpoint $T$, where no subsequent data remain. When $J$ is nonsingular, the center and width of this likelihood, projected onto the mechanical coordinates, are
\begin{equation}
 \bm b_0=\Pi J^{-1}\bm h,\qquad
 V^{(\ell)}=\Pi J^{-1}\Pi^{\mathsf T}.
 \label{eq:original_center}
\end{equation}
We call $\bm b_0$ the future-only likelihood center: it depends on the future record alone and is not a conditional mean formed with a state prior. The superscript $\ell$ labels the future likelihood~\cite{Gammelmark2013}; $V^{(\ell)}$ is its width, not a posterior state covariance. For $j=q,p$, the conventional rule recovers $V_{jj}$ exactly when $V^{(\ell)}_{jj}=V_{jj}$.

\textit{Smoothed trajectory.} The smoothed trajectory combines both parts of the record. The Markov property of the augmented state gives
\begin{equation}
 p(\bm z_t\mid\mathcal Y_-,\mathcal Y_+)
 \propto p(\mathcal Y_+\mid\bm z_t)\,
             p(\bm z_t\mid\mathcal Y_-).
 \label{eq:smoothing_factorization}
\end{equation}
Adding the forward Gaussian coefficients $(P^{-1},P^{-1}\bm f_z)$ to the future coefficients $(J,\bm h)$ and completing the square gives the smoothing covariance and mean~\cite{Mayne1966},
\begin{align}
 P_s&=(P^{-1}+J)^{-1},\notag\\
 \bm s_z&=P_s(P^{-1}\bm f_z+\bm h),\notag\\
 \bm s&=\Pi\bm s_z,\qquad U=\Pi P_s\Pi^{\mathsf T}.
 \label{eq:augmented_smoother}
\end{align}
In the equivalent classical estimation problem (Sec.~\ref{sec:theory}\,A), the smoother residual $\bm x_t-\bm s$, the state minus its conditional mean given the complete record, is uncorrelated with every trajectory derived from the record; the reconstruction identity rests on this orthogonality. The matrix $U$ is the covariance of this residual and is not interpreted as a physical quantum-state covariance~\cite{Tsang2022}. Both $\bm b_0$ and $\bm s$ are calculated in the augmented state and then projected, so they include the correlations between the mechanical and auxiliary variables.

\textit{Past-conditioned center.} The mechanical future-only likelihood integrates over the auxiliary force variables at time $t$ with a flat measure; that is, it treats all their values as equally likely (Appendix~\ref{sec:mechanical_projection}). This can make it very broad (Sec.~\ref{sec:colored_bias_mechanism}). The past record, however, does constrain these variables. Using that information gives the future likelihood of the mechanical state conditioned on the past record,
\begin{align}
 p(\mathcal Y_+\mid\bm x,\mathcal Y_-)
 &=\int p(\mathcal Y_+\mid\bm x,\bm\eta)
       p(\bm\eta\mid\bm x,\mathcal Y_-)\,d\bm\eta\notag\\
 &\propto\frac{p(\bm x\mid\mathcal Y_-,\mathcal Y_+)}
                  {p(\bm x\mid\mathcal Y_-)}.
 \label{eq:main_conditional_likelihood}
\end{align}
For Gaussian forward and smoothed distributions, its precision (inverse width) matrix is $U^{-1}-V^{-1}$. Let $D$ denote the covariance reduction supplied by the future record. If it is positive definite, the center of this likelihood is
\begin{equation}
 \bm b_c=\bm f+M(\bm s-\bm f),\qquad
 M=VD^{-1},\qquad D=V-U\succ0.
 \label{eq:main_conditional_center}
\end{equation}
We call $\bm b_c$ the past-conditioned center. It is a linear combination of the forward and smoothed trajectories, with a gain $M$ calculated from the model, and therefore depends on both parts of the record. When the mechanical coordinates constitute the complete Markov state, $\bm b_c=\bm b_0$. Appendix~\ref{sec:conditional_center_rank} specifies when the future record reduces the uncertainty in every mechanical direction, so that $\bm b_c$ is uniquely defined.

\subsection{The identity and the conventional prescription}

To relate $V$ to trajectories obtained from the measurement record, we first derive an identity for their ensemble covariances, taken over realizations of the record. Finite-record estimates of these covariances are introduced in Sec.~\ref{sec:experiment}\,C. Define the covariances of the trajectory differences by
\begin{align}
 C_{fb}&=\operatorname{Cov}(\bm f-\bm b_0),\notag\\
 C_{sf}&=\operatorname{Cov}(\bm s-\bm f),\notag\\
 C_{sb}&=\operatorname{Cov}(\bm s-\bm b_0).
 \label{eq:difference_covariances}
\end{align}
The Markov property of the augmented state and the smoothing orthogonality (Sec.~\ref{sec:future_information}) imply
\begin{align}
 C_{fb}&=V+V^{(\ell)},\notag\\
 C_{sf}&=V-U,\qquad C_{sb}=V^{(\ell)}-U.
 \label{eq:difference_covariance_values}
\end{align}
The first relation adds the covariances of the forward error and the future-likelihood error; the two are independent because the latter contains only noise after $t$. The second and third follow from the orthogonality and give the uncertainty reduction achieved by smoothing, relative to the forward trajectory and to the future-only likelihood center; their difference $C_{sf}-C_{sb}$ is the imbalance $V-V^{(\ell)}$. Combining the three relations cancels the future-likelihood width and the smoothing covariance:
\begin{equation}
 V=\tfrac12(C_{fb}+C_{sf}-C_{sb}).
 \label{eq:three_identity}
\end{equation}
This identity defines the three-trajectory covariance estimator, which reconstructs all three independent components from the trajectory differences. Within the specified linear-Gaussian model, the identity holds at the ensemble level for the exact conditional quantities at each time. Appendix~\ref{app:identity} gives the proof, the conditions for existence of the future-only likelihood center, and an equivalent cross-covariance form.

Equation~\eqref{eq:three_identity} also holds unchanged when $\bm b_0$ is replaced by $\bm b_c$ in all three difference covariances: smoothing orthogonality gives $\operatorname{Cov}(\bm f-\bm s)=D$ and $MD=DM^{\mathsf T}=V$ (Appendix~\ref{sec:conditional_center_rank}). We call this variant the past-conditioned reconstruction. Because $\bm b_c$ is formed from $\bm f$ and $\bm s$, the record enters it only through the covariance of $\bm a=\bm f-\bm s$; the gain $M$ is taken from the model.

Rossi \textit{et al.}~\cite{Rossi2019} approximated the sum of the forward and retrodictive variances by twice the forward variance in their experimental regime. For comparison, we apply the corresponding half-difference rule to the future-only likelihood center $\bm b_0$ defined here and use a separate rule for the off-diagonal entry. We call this combination the conventional prescription and denote its full-band ensemble prediction by $V_2$. The subscript 2 labels this two-trajectory prescription:

\begin{align}
 (V_2)_{qq}&=\tfrac12\operatorname{Var}(f_q-b_{0,q}),\notag\\
 (V_2)_{pp}&=\tfrac12\operatorname{Var}(f_p-b_{0,p}),\notag\\
 (V_2)_{qp}&=-\operatorname{Cov}(f_q,f_p).
 \label{eq:conventional_rule}
\end{align}

For each diagonal entry, the bias is half the difference between the corresponding future-likelihood and forward variances [Eq.~\eqref{eq:difference_covariance_values}]. Thus, a broad future likelihood can produce a positive bias, which the three-trajectory identity cancels.
The off-diagonal entry uses the forward-trajectory prescription
of Ref.~\cite{Hatakeyama2026}. For stationary full-band motion with $\dot q=\omega p$, where $\omega$ is the mechanical angular frequency, and no direct position diffusion, the unconditional position--momentum covariance vanishes. The law of total covariance therefore gives $-\operatorname{Cov}(f_q,f_p)=V_{qp}$ (Appendix~\ref{app:colored_bias}).
By contrast, Eq.~\eqref{eq:three_identity} applies to all three entries without these additional conditions. It keeps its matrix form under linear coordinate transformations and extends to multiple modes whenever the required trajectory centers exist (Appendix~\ref{sec:identity_rank_endpoint}).

\Needspace{7\baselineskip}
\subsection{Origin of the bias with colored force noise}
\label{sec:colored_bias_mechanism}

For the oscillator model considered below, the conventional prescription overestimates the position variance, and each observable colored-noise component adds to the bias however weakly it couples. We show this analytically for a finite OU representation of the colored force, using the future-only likelihood center of Sec.~\ref{sec:future_information}.

Consider a stationary continuous oscillator with
mechanical angular frequency $\omega>0$, $\dot q=\omega p$,
momentum damping rate $\gamma>0$, and force noise
acting on momentum. Represent its colored force by $N_{\rm OU}$ OU variables with distinct decay rates $a_j>0$ and nonzero couplings, so that each is observable from the future output. Normalize the readout to
$dY=q\,dt+dw$ and denote its white-noise intensity by
$\mathbb E[dw^2]=R_y\,dt$, so $\mathcal R=R_y$ in this normalization. The OU driving noises are mutually independent and independent of the direct white-force and readout noise, which may be correlated with each other. Appendix~\ref{app:colored_bias} states the precise assumptions and gives
\begin{equation}
 (V_2)_{qq}-V_{qq}
   =R_y\left(\gamma+\sum_{j=1}^{N_{\rm OU}}a_j\right).
 \label{eq:colored_bias_qq}
\end{equation}
Each observable OU component adds $R_y a_j$ to this full-band bias, whatever its nonzero coupling strength. This term is the noise variance of the readout averaged over the decay time $a_j^{-1}$ of that component, so a faster decay gives a larger term.

Each component contributes because the future-only likelihood does not know the amplitude of each OU variable at the reconstruction time. Although the variable loses memory over a time $a_j^{-1}$, the future record must distinguish its short-lived force transient from the initial mechanical state. Integrating over these unknown amplitudes broadens the mechanical likelihood, even for rapidly decaying OU modes. This is the origin of $\sum_j a_j$ in Eq.~\eqref{eq:colored_bias_qq}. The corresponding momentum bias in Appendix~\ref{app:colored_bias}\,4 [Eq.~\eqref{eq:cb_bias}] includes the force--readout noise correlation and can be much larger.

\subsection{What a finite-band comparison tests}
\label{sec:band_comparison}

A band-limited comparison does not test $V$ directly. Equal-time covariances of real stationary processes are the integrals of the real parts of their cross spectra over positive frequencies, and restricting these integrals to a band can shift the result away from $V$, even when the three-trajectory identity is exact. We therefore define the finite-band prediction, with which a band-limited reconstruction should be compared, and a relative error that measures how far this prediction is from $V$. Here ``full band'' includes all frequencies represented by the specified model.

Let $S_{fb}(\nu)$, $S_{sf}(\nu)$, and $S_{sb}(\nu)$ be the one-sided matrix spectra of the corresponding differences at frequency $\nu$ in hertz. Their signed combination is
\begin{equation}
 S_3(\nu)=\tfrac12[S_{fb}(\nu)+S_{sf}(\nu)-S_{sb}(\nu)].
 \label{eq:three_spectrum}
\end{equation}
An equivalent cross-spectrum expression is given in Appendix~\ref{app:identity}\,6 [Eq.~\eqref{eq:app_cross_spectrum}]. For a stationary continuous model with stationary filters derived from that model, the full-band and band-restricted integrals are
\begin{align}
 V&=\int_0^{\infty}\operatorname{Re}S_3(\nu)\,d\nu,\notag\\
 P_{3,B}&=\int_B\operatorname{Re}S_3(\nu)\,d\nu,
 \label{eq:theory_band_integral}
\end{align}
where $B$ is the analysis band. The signed spectral integral $P_{3,B}$ is the finite-band prediction for the reconstruction; it equals $V$ in the full-band limit. In the experiment, the analysis band is narrower than the model's full band because detector response, technical noise, and the validated model range limit the comparison.

The band-truncation error is the largest relative difference between $P_{3,B}$ and $V$ over all quadrature directions,
\begin{equation}
 \begin{aligned}
 \epsilon_B&=\left\|V^{-1/2}(P_{3,B}-V)V^{-1/2}\right\|_2\\
 &=\sup_{\bm e\ne0}
 \frac{|\bm e^{\mathsf T}(P_{3,B}-V)\bm e|}
      {\bm e^{\mathsf T}V\bm e}.
 \end{aligned}
 \label{eq:band_accuracy}
\end{equation}
Here $\|\cdot\|_2$ is the spectral norm and $\bm e$ is any nonzero real vector selecting a quadrature direction. For example, $\epsilon_B=0.1$ means that the finite-band prediction is within $10\%$ of $V$ in every quadrature direction, including rotated ones: $\epsilon_B\leq\epsilon$ is equivalent to $(1-\epsilon)V\preceq P_{3,B}\preceq(1+\epsilon)V$, where $\preceq$ denotes positive-semidefinite order. The same definition applies with either $\bm b_0$ or $\bm b_c$.

A finite-band prediction, or its estimate from a finite record, can represent a quantum-state covariance only if it is positive definite and, for the single mode and normalization used here, has a determinant of at least $1/4$. For the prediction, $V\succ0$ and $\epsilon_B<1$ guarantee that $P_{3,B}$ is positive definite.

We compare the finite-record reconstruction with its finite-band prediction $P_{3,B}$. Whether the reconstruction can also serve as an estimate of $V$ depends on whether the finite-band offset $P_{3,B}-V$ and the propagated uncertainty are small compared with the required accuracy. The error $\epsilon_B$ is defined at fixed model parameters. The experimental analysis varies the measured inputs within their uncertainties and evaluates the scatter due to the finite record (Sec.~\ref{sec:results}\,A).

\Needspace{12\baselineskip}
\subsection[Bandwidth bounds and finite-horizon effects]{Bandwidth bounds and finite-horizon effects}
\label{sec:ideal_bandwidth}

With the past-conditioned center $\bm b_c$, we formulate a bandwidth criterion in terms of the covariance of the trajectory difference $\bm a=\bm f-\bm s$. The criterion accounts for the variance omitted in each direction and its amplification in the reconstruction.

For a stationary model with fixed filters derived from it, $\bm a$ has covariance $D\succ0$ and a one-sided matrix autospectrum $S_{aa}$, which is positive semidefinite and integrates to $D$ over the full band. Equations~\eqref{eq:main_conditional_center} and \eqref{eq:three_identity} give
\begin{align}
 D_B&=\int_B\operatorname{Re}S_{aa}(\nu)\,d\nu,\notag\\
 P_{3c,B}&=\operatorname{sym}(D_BD^{-1}V),
 \label{eq:conditional_band_prediction}
\end{align}
where $D_B$ is the part of $D$ contained in $B$ and $\operatorname{sym}X=(X+X^{\mathsf T})/2$. The finite-band prediction thus differs from $V$ only through the part of $D$ that the band omits.

Two numbers quantify this omission:
\begin{align}
 \zeta_B&=\left\|D^{-1/2}(D-D_B)D^{-1/2}\right\|_2,\notag\\
 g_c&=\sqrt{\operatorname{cond}_2
                    (V^{-1/2}DV^{-1/2})}.
 \label{eq:band_capture_metric}
\end{align}
Here $\zeta_B$ is the largest fraction of the variance of $\bm a$ that the band omits in any direction, and $g_c\geq1$ bounds how much this omission is amplified in the reconstruction; $\operatorname{cond}_2X=\|X\|_2\|X^{-1}\|_2$ is the condition number. For any band $B$, Appendix~\ref{app:bandwidth} proves
\begin{equation}
 \zeta_B\leq\epsilon_{B,c}\leq g_c\zeta_B,
 \qquad 0\leq\zeta_B\leq1,
 \label{eq:conditional_band_bound}
\end{equation}
where $\epsilon_{B,c}$ is Eq.~\eqref{eq:band_accuracy} evaluated with $\bm b_c$. A band therefore meets a tolerance $\epsilon$ if $\zeta_B\leq\epsilon/g_c$, and cannot meet it unless $\zeta_B\leq\epsilon$. Because $S_{aa}$ is positive semidefinite, widening the band never increases $\zeta_B$, although the error itself can vary nonmonotonically.

This bound applies to $\bm b_c$, not to $\bm b_0$. The two centers differ by a residual, $\bm r_\perp=(\bm f-\bm b_0)-M\bm a$, whose cross covariance with $\bm a$ vanishes over the full band but can be nonzero over a restricted band. For a particular band, either center can give the smaller error (Appendix~\ref{app:bandwidth}\,1), but the error with $\bm b_0$ is not bounded in this way and can be much larger. In the ideal model of Appendix~\ref{sec:sm_ideal_comparison}, the band $[0.30,40]\nu_0$ around the static-spring frequency $\nu_0$ gives an error of $7.71\%$ with $\bm b_c$ and more than $400\%$ with $\bm b_0$.

The duration of the future record sets a separate information limit. Extending that record at fixed $t$ adds information while $V$ stays fixed, so the smoothing-error variance decreases or stays constant in every quadrature direction. For the stable models considered here, once the forward filter has reached stationarity, $U$ and, where defined, $V^{(\ell)}$ approach their stationary limits as the future record is extended (Appendix~\ref{sec:identity_rank_endpoint}). The simulations of Sec.~\ref{sec:results}\,A assess the finite-record processing and endpoint effects of the implemented stationary-gain filters. SM Sec.~\ref{sec:sm_future_information} compares the finite-horizon smoothing covariance with its stationary limit.

\section{Optomechanical experiment and analysis}
\label{sec:experiment}

We connect the measured inputs to a physical model, then define the calibration and processing used in the covariance comparisons.

\subsection{Apparatus and physical model}

\begin{figure*}[t]
\centering
\includegraphics[width=\textwidth]{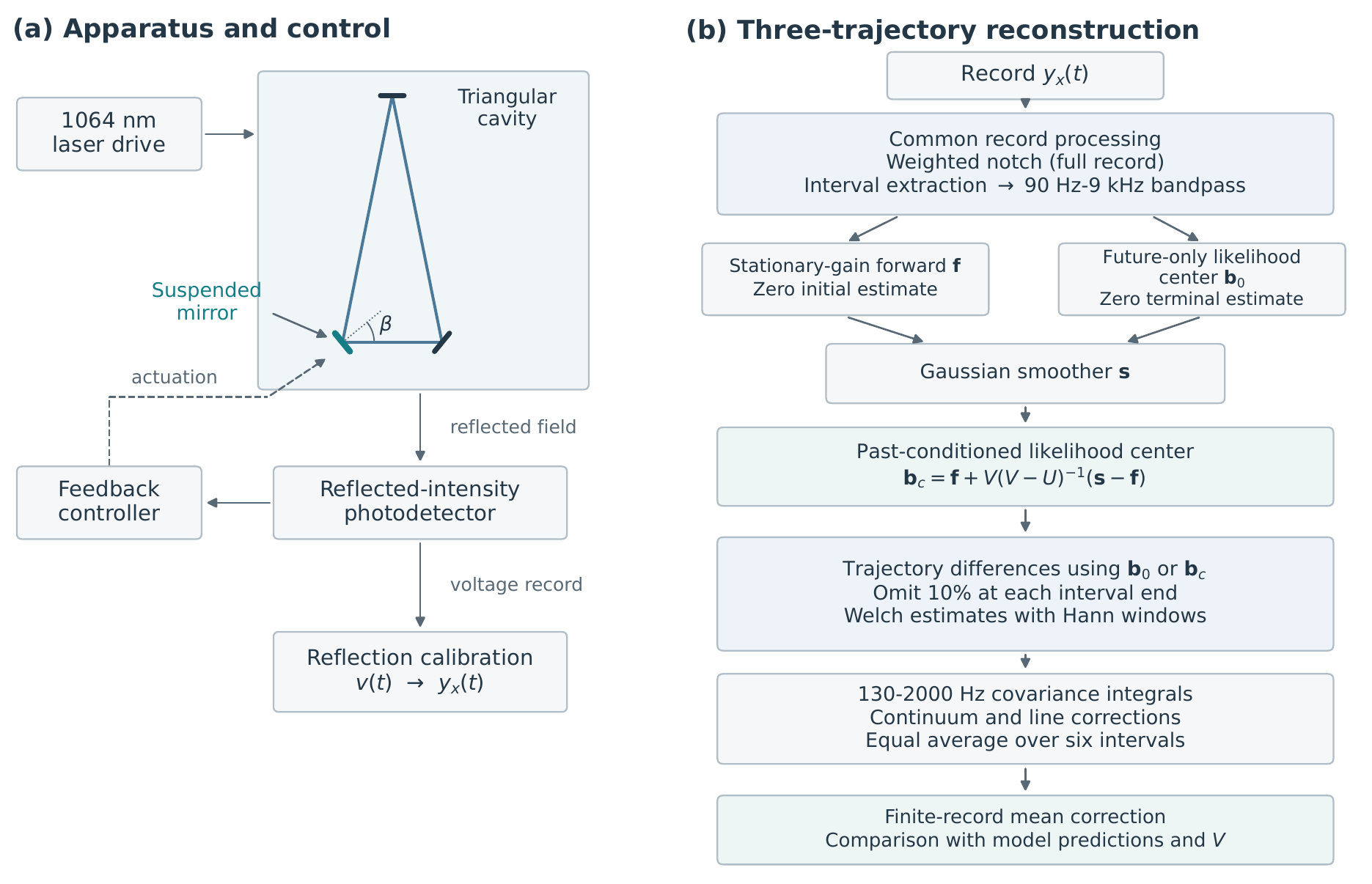}
\caption{Experimental setup and covariance reconstruction.
(a) The photodiode records the total reflected intensity from the triangular cavity containing the suspended mirror. The dashed line shows the effective feedback-cooling path. The lower-left mirror is the silica-fiber-suspended test mirror; $\beta$ denotes the incidence angle.
(b) The same preprocessing, stationary filters, and spectral integration are applied to the experimental and simulated records. The forward trajectory, future-only likelihood center, and smoothed trajectory give the reconstruction statistics. The past-conditioned center is formed from the forward and smoothed trajectories. Section~\ref{sec:experiment}\,C defines the processing and corrections.}
\label{fig:apparatus_estimator}
\end{figure*}

To test the reconstruction, we reanalyze the detector record of the suspended-mirror cavity in Ref.~\cite{Matsumoto2019} and build its physical model. The apparatus is shown in Fig.~\ref{fig:apparatus_estimator}(a).
A triangular ring cavity driven at
$1064\,\mathrm{nm}$ contains a $7.71\,\mathrm{mg}$ mirror suspended from a
silica fiber. A detuned field with total incident power before mode matching
$P_{\rm in}=30\,\mathrm{mW}$ acts as an optical spring: its radiation pressure depends on the mirror position and shifts the mechanical
resonance from $\nu_{\rm bare}=4.527\,\mathrm{Hz}$ to approximately
$283.5\,\mathrm{Hz}$, which we use as the reference frequency for calibration and quadrature normalization. Only the fraction of incident power coupled into the cavity mode drives the optical spring (SM Sec.~\ref{sec:sm_calibration_dependencies}). At the chosen detuning, the optical spring provides a restoring force and the optical field produces negative damping. Feedback from the reflected-intensity readout damps the motion near the trapped resonance [Fig.~\ref{fig:apparatus_estimator}(a)]. The interval models represent this feedback through the effective damping rate.
The same readout provides the displacement record. Optical dilution reduces the effective mechanical loss of the pendulum mode~\cite{Corbitt2007}. The measured bare-pendulum frequency and intrinsic quality factor $Q_{\rm bare}$ set the thermal-force reference~\cite{Matsumoto2019,GonzalezSaulson1994}.

We analyze six selected portions of the record, called verification intervals, in which the resonance frequency is comparatively stable. The index $i=1,\ldots,6$ labels these intervals. Their waveform-derived resonance frequencies are $\nu_i$, their angular frequencies are $\omega_i=2\pi \nu_i$, and the local models use the same effective quality factor $Q_{\rm sh}=1000$ for all six intervals. This value was retained from the preliminary model screening and held fixed during uncertainty propagation. Section~\ref{sec:experiment}\,B describes the interval selection and frequency estimation.

We derive the reduced mechanical and readout equations~\eqref{eq:exp_reduced} from the cavity dynamics, then specify their noise sources and normalization.

\textit{Optical model.} To keep the correlation between force and readout noise, we derive both from the same optical inputs. Let $m$ denote the mirror mass, $x_m$ its physical displacement, and $p_{\rm dim}$ its dimensionful momentum. With cavity-mode annihilation operator $\hat a$, the driven optomechanical Hamiltonian is

\begin{align}
 H_{\rm om}(t)={}&\frac{p_{\rm dim}^2}{2m}
 +\frac{m\omega_{\rm bare}^2x_m^2}{2}
 +\hbar(\omega_{\mathrm{cav}}+Gx_m)\hat a^\dagger\hat a\nonumber\\
 &+\mathrm i\hbar\bigl[\mathcal E(t)\hat a^\dagger e^{-\mathrm i\phi_L(t)}
 -\mathcal E(t)^*\hat a e^{\mathrm i\phi_L(t)}\bigr],
 \label{eq:exp_hamiltonian}
\end{align}
where $\hbar$ is the reduced Planck constant, $\omega_{\rm bare}=2\pi \nu_{\rm bare}$, $\omega_{\mathrm{cav}}$ and $\omega_L$ are the cavity and mean laser angular frequencies, and $G=\partial\omega_{\mathrm{cav}}/\partial x_m>0$. The coherent cavity-drive amplitude is $\mathcal E(t)$, with steady value $\mathcal E_0$. The classical incident fractional-power fluctuation $u(t)$ enters through $\mathcal E(t)=\mathcal E_0[1+u(t)/2]$ to first order. The drive phase obeys $\dot\phi_L=\omega_L+\delta\omega_L(t)$, where $\delta\omega_L(t)$ is the fluctuation of the laser angular frequency. Frequency noise is thus included in the cavity equations, before the cavity dynamics are eliminated.

In the instantaneous laser frame, set $\Delta=\omega_L-\omega_{\mathrm{cav}}$ and absorb the static displacement into $\omega_{\mathrm{cav}}$. Write $\hat a=\bar a+\delta\hat a$ and choose the phase so that $\bar a=\mathcal E_0/(\kappa-\mathrm i\Delta)=\sqrt{n_{\mathrm{cav}}}$ is real, where $n_{\mathrm{cav}}$ is the mean intracavity photon number. The angular cavity half width is $\kappa=\kappa_{\rm in}+\kappa_{\rm loss}$, with input coupling $\kappa_{\rm in}$ and remaining optical loss $\kappa_{\rm loss}$. With $\mathcal G=G\sqrt{n_{\mathrm{cav}}}$, $X_{\mathrm{cav}}=\delta\hat a+\delta\hat a^\dagger$, and $Y_{\mathrm{cav}}=(\delta\hat a-\delta\hat a^\dagger)/\mathrm i$, the linearized optical equations are

\begin{align}
 \dot X_{\mathrm{cav}}={}&-\kappa X_{\mathrm{cav}}-\Delta Y_{\mathrm{cav}}+\kappa\sqrt{n_{\mathrm{cav}}}\,u\nonumber\\
       &+\sum_{j_{\rm port}\in\{\mathrm{in},\mathrm{loss}\}}\sqrt{2\kappa_{j_{\rm port}}}\,X_{j_{\rm port}},\nonumber\\
 \dot Y_{\mathrm{cav}}={}&-\kappa Y_{\mathrm{cav}}+\Delta X_{\mathrm{cav}}-2\mathcal Gx_m\nonumber\\
       &+2\sqrt{n_{\mathrm{cav}}}\,\delta\omega_L-\Delta\sqrt{n_{\mathrm{cav}}}\,u\nonumber\\
 &+\sum_{j_{\rm port}\in\{\mathrm{in},\mathrm{loss}\}}\sqrt{2\kappa_{j_{\rm port}}}\,Y_{j_{\rm port}}.
 \label{eq:exp_optical_langevin}
\end{align}

Here $j_{\rm port}\in\{\mathrm{in},\mathrm{loss}\}$ labels the input and loss ports. Their vacuum quadratures $X_{j_{\rm port}},Y_{j_{\rm port}}$ are independent, with symmetrized white-noise correlations normalized to a delta function in time. The radiation-pressure fluctuation is $-\hbar\mathcal G X_{\mathrm{cav}}$.

\textit{Detection.} The photodiode collects the cavity-matched reflected mode ($\parallel$) and the orthogonal, promptly reflected light ($\perp$). The matched mode obeys the cavity input--output relation. For $j\in\{\parallel,\perp\}$, the detected fields and photocurrent $I$ satisfy
\begin{align}
 \hat a_{{\rm r},\parallel}&=\hat a_{{\rm in},\parallel}-\sqrt{2\kappa_{\rm in}}\,\hat a,\nonumber\\
 \hat a_{{\rm det},j}&=\sqrt{\eta_{\rm PD}}\,\hat a_{{\rm r},j}
                 +\sqrt{1-\eta_{\rm PD}}\,\hat v_j,\nonumber\\
 I&\propto\sum_j\hat a_{{\rm det},j}^{\dagger}\hat a_{{\rm det},j}.
 \label{eq:exp_detected_output}
\end{align}
Here $\eta_{\rm PD}$ is the photodiode quantum efficiency and the $\hat v_j$ are independent detection-loss vacuum fields. Both modes contribute to the detected power. Only the matched mode responds to mirror displacement. A separate pickoff before the cavity measures incident relative intensity noise (RIN). The reflection calibration gives the displacement-equivalent readout $y_x(t)=\alpha v(t)$ from the detector voltage $v(t)$. Equation~\eqref{eq:exp_units} defines its normalization.

\textit{Reduced model.} The cavity half width $\kappa/(2\pi)\simeq0.83\,\mathrm{MHz}$ greatly exceeds the analysis frequencies. Adiabatic elimination of Eq.~\eqref{eq:exp_optical_langevin} gives

\begin{align}
 X_{\mathrm{cav}}&=\sqrt{n_{\mathrm{cav}}}\,u+
 \frac{2\Delta\mathcal G(x_m-n_{\rm fn})+N_X}{\kappa^2+\Delta^2},\nonumber\\
 n_{\rm fn}&=\frac{\delta\omega_L}{G}=\frac{2\pi}{G}\delta\nu_L.
 \label{eq:exp_adiabatic_amplitude}
\end{align}

Here $\delta\nu_L$ is the laser-frequency fluctuation in hertz, and $n_{\rm fn}$ is its equivalent displacement. The term $N_X=\sum_{j_{\rm port}}\sqrt{2\kappa_{j_{\rm port}}}(\kappa X_{j_{\rm port}}-\Delta Y_{j_{\rm port}})$ contains the optical vacuum inputs. The optical stiffness is

\begin{equation}
 k_{\rm opt}=\frac{2\hbar G^2n_{\mathrm{cav}}}{\kappa}
                  \frac{\delta}{1+\delta^2},\qquad \delta=\Delta/\kappa.
 \label{eq:exp_spring}
\end{equation}

The positive detuning gives a restoring optical spring. The measured resonance $\omega_i$ sets the dynamical stiffness $m\omega_i^2$ and effective damping $\gamma_i=\omega_i/Q_{\rm sh}$ in interval $i$. We infer the detuning from $\omega_i$ using SM Eq.~\textup{(\ref{eq:sm_spring_graph})} and evaluate the frequency-noise force coefficient $k_{{\rm opt},i}$ from the optical calibration [Eq.~\eqref{eq:exp_spring} and SM Sec.~\ref{sec:frequency_noise_limits}].

The reduced equations, together with the OU force dynamics below, are
\begin{align}
 \dot x_m&=p_{\rm dim}/m,\nonumber\\
 \dot p_{\rm dim}&=-m\omega_i^2x_m-\gamma_i p_{\rm dim}
             +F_{\rm th}+F_u u\nonumber\\
             &\quad+k_{{\rm opt},i}n_{\rm fn}+F_{\rm vac},\nonumber\\
 y_x&=x_m+X_u u-n_{\rm fn}+v_{\rm vac}.
 \label{eq:exp_reduced}
\end{align}
Here $F_{\rm vac}$ and $v_{\rm vac}$ are the correlated force and readout combinations of the vacuum quadratures. For the fractional incident-power fluctuation $u$, the force coefficient is $F_u=-\hbar G n_{\mathrm{cav}}$, and $X_u$ is the direct response of the displacement-referred readout to the same fluctuation. It equals the total collected reflected power divided by its displacement slope (SM Sec.~\ref{sec:sm_optical_coefficients}). The intensity force and direct-readout coefficients have the same sign. Frequency noise has opposite relative signs: force $+k_{{\rm opt},i}n_{\rm fn}$ and direct readout $-n_{\rm fn}$. Both paths enter its contribution to the detected spectrum. The reduced model assumes linear response with small fluctuations. The simulations assume Gaussian driving noise (SM Sec.~\ref{sec:sm_empirical_uncertainty}).

\textit{Classical optical noise.} The classical inputs $u$ and $n_{\rm fn}$ and the thermal OU drivers are mutually independent; each optical input has its own force--readout correlation. We denote the measured RIN amplitude spectral density (ASD), divided by the shot-noise ASD, by $R_{\rm RIN}$. The classical excess PSD of $u$, referred to the total incident power $P_{\rm in}=30\,\mathrm{mW}$ before mode matching, and the PSD of $n_{\rm fn}$ are
\begin{align}
 S_u&=(R_{\rm RIN}^2-1)\frac{2\hbar\omega_L}{P_{\rm in}},\nonumber\\
 S_{n_{\rm fn}}(\nu)&=\left(\frac{2\pi}{G}\right)^2 S_{\delta\nu_L}(\nu).
 \label{eq:exp_classical_noise_psds}
\end{align}
The vacuum inputs of the cavity ports, unmatched reflected light, and detection loss keep their quantum normalization.

For Fourier components proportional to $e^{-2\pi\mathrm i\nu t}$, the mechanical susceptibility $\chi_i$ and the displacement-referred output contributions from intensity and frequency noise are
\begin{align}
 \chi_i(\nu)&=\frac{1}{m[\omega_i^2-(2\pi\nu)^2-\mathrm i\gamma_i(2\pi\nu)]},\nonumber\\
 S_{y_x,\mathrm{int}}(\nu)&=|X_u+\chi_i(\nu)F_u|^2S_u,\nonumber\\
 S_{y_x,\mathrm{fn}}(\nu)&=|k_{{\rm opt},i}\chi_i(\nu)-1|^2S_{n_{\rm fn}}.
 \label{eq:exp_intensity_interference}
\end{align}
Below the resonance, the two frequency-noise paths partly cancel in the readout. For a resonance at $283.5\,\mathrm{Hz}$, the frequency-noise contribution to the output PSD is about $0.04$ times its contribution to the mirror-displacement PSD at $130\,\mathrm{Hz}$ and $0.25$ times at $200\,\mathrm{Hz}$, using the spectral ratio in SM Sec.~\ref{sec:frequency_noise_limits}.

\textit{Thermal force.} The colored thermal force $F_{\rm th}$ is represented by six independent
OU variables $\eta_j$,
\begin{equation}
 F_{\rm th}=\sum_{j=1}^{6}F_j\eta_j,\qquad
 \dot \eta_j=-a_j\eta_j+\sqrt{2a_j}\,\xi_j^{\rm OU},
 \label{eq:exp_ou}
\end{equation}
where $a_j$ set the force-memory times $a_j^{-1}$ and $F_j$ convert the dimensionless
OU variables into force.
Each $\eta_j$ has unit stationary variance. The independent white inputs satisfy $\mathbb E[\xi_j^{\rm OU}(t)\xi_l^{\rm OU}(t')]=\delta_{jl}\delta(t-t')$. The rates and coefficients are chosen to approximate the structural-damping force spectrum
\begin{equation}
 S_{FF,\rm th}(\nu)\simeq
 R_F\frac{4k_{\rm B}T_bm(2\pi \nu_{\rm bare})^2}{Q_{\rm bare}(2\pi\nu)},\qquad \nu>0.
 \label{eq:exp_force}
\end{equation}
Here $k_{\rm B}$ is Boltzmann's constant and $T_b=300\,\mathrm K$ is the bath temperature. We use a fixed thermal-force prefactor $R_F=4$ throughout the analysis band. Its value follows the fourfold reduction of the optically diluted quality factor reported for this suspension near $280\,\mathrm{Hz}$~\cite{Matsumoto2019} and attributed to pendulum--pitch mixing~\cite{Sugiyama2023}. We normalize the six-OU representation so that its total force-noise power over 130--2000 Hz agrees with Eq.~\eqref{eq:exp_force}, the fluctuation--dissipation relation linking thermal fluctuations to suspension loss.
\label{sec:ou_optimization}

\textit{Laser-frequency noise.} For the frequency-noise term in Eq.~\eqref{eq:exp_intensity_interference}, we assume a free-running laser frequency-noise ASD of
\[
 \frac{7.66\times10^3}{\nu/\mathrm{Hz}}\,
 \frac{\mathrm{Hz}}{\sqrt{\mathrm{Hz}}}.
\]
With this input, the measured detector-output PSD is $3$--$13$ times the ideal high-gain servo prediction in all 20 frequency bins centered above $330\,\mathrm{Hz}$ [SM Fig.~\ref{fig:sm_ideal_servo_budget}(b)]. We therefore constrain the noise suppression using the open-loop gain measured with the servo's additional gain stage (boost) enabled (SM Fig.~\ref{fig:sm_servo_response}). The measured output spectrum and preliminary covariance comparisons guide the choice of filter. In propagating uncertainties, we vary its poles and zeros with the measured servo response while keeping the filter structure and free-running spectrum fixed (SM Sec.~\ref{sec:sm_noise_scope}). Table~\ref{tab:main_parameters} lists the nominal input values used in the reference calculations.

\textit{Normalization.} The target quadratures are $\bm x=(q,p)^{\mathsf T}$, with
\begin{equation}
 \begin{aligned}
 q&=\frac{x_m}{\sqrt{\hbar/(m\omega_i)}},\qquad
 p=\frac{p_{\rm dim}}{\sqrt{\hbar m\omega_i}},\\
 y&=\frac{y_x}{\sqrt{\hbar/(m\omega_i)}}.
 \end{aligned}
 \label{eq:exp_units}
\end{equation}
This is the normalization introduced in Sec.~\ref{sec:theory}. In the continuous-time notation of Eq.~\eqref{eq:theory_state}, $Y$ is the time integral of the normalized detector output $y$. Because this normalization depends on $m$ and $\omega_i$, we convert estimates and predictions to common reference quadratures defined using the nominal mass $7.71\,\mathrm{mg}$ and reference frequency $283.5\,\mathrm{Hz}$ before averaging over intervals. Variations of the measured inputs and calibration within their uncertainties are propagated jointly through the reconstruction and its prediction (SM Sec.~\ref{sec:sm_calibration_dependencies}).

\begin{table}[t]

\caption{Nominal inputs of the adopted physical model. The effective quality factor is fixed across all six intervals. The listed $R_{\rm RIN}=6.12$ is the nominal calculation value. Uncertainty propagation assumes $R_{\rm RIN}\sim\mathrm{Uniform}(4.30,6.12)$. SM Secs.~\ref{sec:sm_calibration_dependencies} and \ref{sec:sm_noise_scope} specify the calibration dependencies and noise assumptions.}
\label{tab:main_parameters}
\begin{ruledtabular}
\begin{tabular}{lr}
Quantity & Nominal value \\
Mirror mass $m$ & $7.71\,\mathrm{mg}$ \\
Bare frequency $\nu_{\rm bare}$ & $4.527\,\mathrm{Hz}$ \\
Intrinsic quality factor $Q_{\rm bare}$ & $9.9\times10^4$ \\
Bath temperature $T_b$ & $300\,\mathrm K$ \\
Cavity half width $\kappa/(2\pi)$ & $834.8\,\mathrm{kHz}$ \\
Incident power $P_{\rm in}$ & $30\,\mathrm{mW}$ \\
Photodiode efficiency $\eta_{\rm PD}$ & $0.85$ \\
Nominal detuning $\delta_0$ & $0.0703$ \\
Displacement scale $\alpha_0$ & $2.15\times10^{-10}\,\mathrm{m/V}$ \\
Shared effective quality factor $Q_{\rm sh}$ & $1000$ \\
Thermal-force prefactor $R_F$ & $4$ \\
Intensity ASD ratio $R_{\rm RIN}$ & $6.12$ \\
\end{tabular}
\end{ruledtabular}

\end{table}

\subsection{Calibration and interval parameters}

To convert the detector record to displacement and determine the parameters of each interval, we use the calibrations underlying Table~\ref{tab:main_parameters}. The cavity scans, optical-spring response, suspension ring-down, and voltage conversion are documented with their uncertainty dependencies in SM Sec.~\ref{sec:sm_auxiliary_calibrations} and Fig.~\ref{fig:sm_auxiliary_calibrations}.
The reflection fit includes the mode-matching efficiency of the scan. On the undercoupled branch ($\rho_{\rm in}=0.22246$), the optical-spring calibration gives an incidence angle $\beta=(0.686\pm0.060)\,\mathrm{rad}$ ($39.3^\circ$), consistent with the design angle of $42.26^\circ$. The nominal detuning $\delta_0$ is the value of $\delta=\Delta/\kappa$ at the reference frequency. Of the two detunings that give the operating resonance frequency, the measured frequency drift supports the smaller one (SM Sec.~\ref{sec:sm_auxiliary_calibrations}).

\begin{figure*}[t]
\centering
\includegraphics[width=\textwidth]{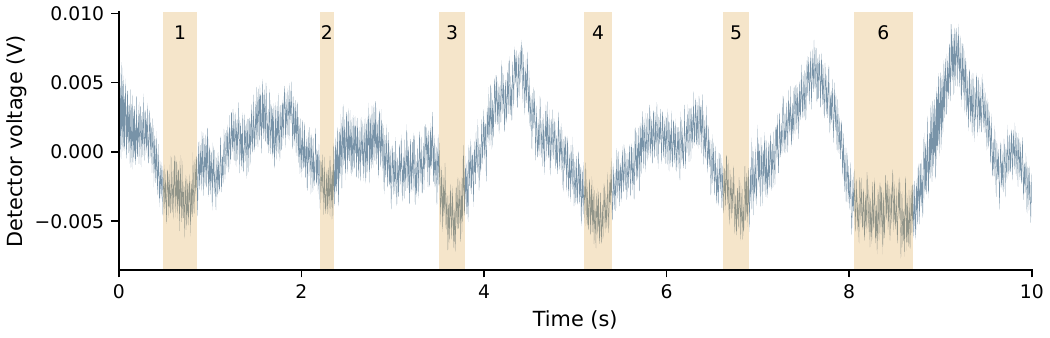}
\caption{First $10$ s of the raw reflected-intensity detector voltage from the suspended-mirror experiment of Ref.~\cite{Matsumoto2019}. The trace shows the minimum and maximum raw voltage in successive 2-ms bins. Shading and numbers mark the six verification intervals; SM Sec.~\ref{sec:sm_stationary_implementation} gives their boundaries, and Fig.~\ref{fig:interval_psd} gives their resonance frequencies.}
\label{fig:raw_record}
\end{figure*}

\begin{figure*}[!t]
\centering
\includegraphics[width=\textwidth]{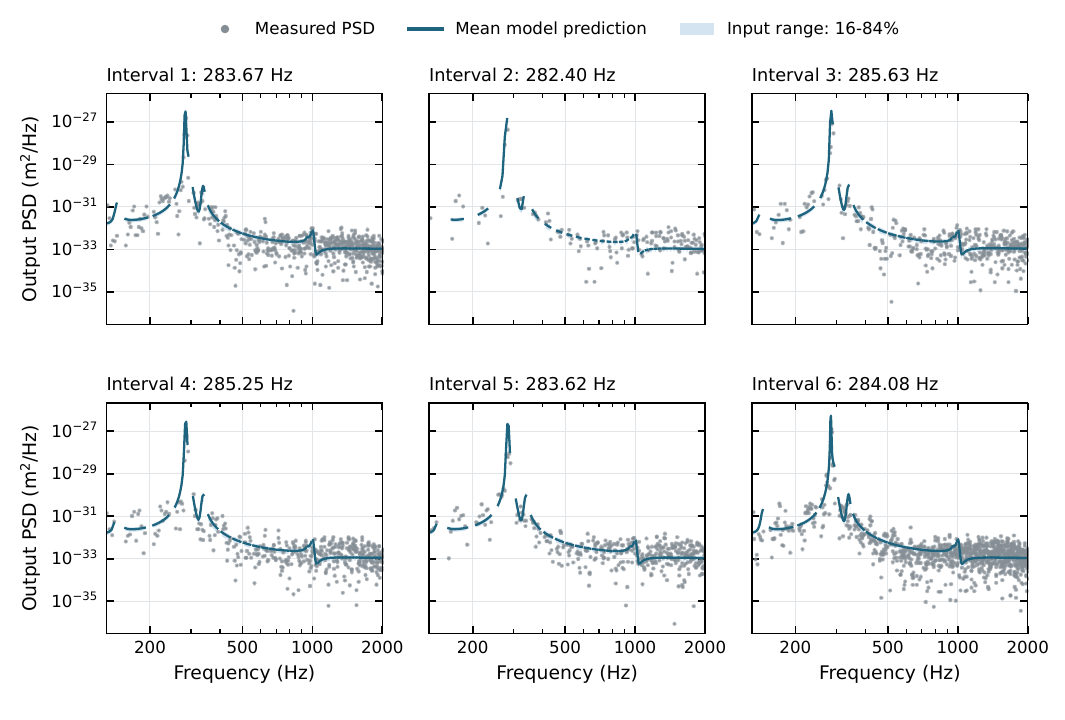}
\caption{One-sided output displacement PSDs per hertz for the six verification intervals. Points show full-interval periodograms after time-domain subtraction of the fitted mains-harmonic waveform; gaps mark omitted harmonic neighborhoods (SM Sec.~\ref{sec:sm_line_processing}). Curves show the mean model prediction over the 64 input conditions, including the finite interval length and the Hann window; shading gives their pointwise 16th--84th percentile range. Spectra use the common nominal displacement calibration and $Q_{\rm sh}=1000$.}
\label{fig:interval_psd}
\end{figure*}

The first $10$ s of the detector record,
used for interval selection and displayed in Fig.~\ref{fig:raw_record},
show slow oscillatory variations; the trapped mechanical resonance also drifts
with time. To reduce the effect of this nonstationarity, we
selected six intervals by exploratory inspection after acquisition. Their boundaries remain fixed in the comparisons between reconstructed and predicted covariances. We estimate each $\nu_i$ from the waveform phase evolution. The frequency-estimation rule was fixed using parts of the record outside these intervals (SM Sec.~\ref{sec:sm_waveform_frequency}).
The resulting frequencies are given in Fig.~\ref{fig:interval_psd}.

We express the detector voltage, sampled at $1\,\mathrm{MHz}$, in nominal displacement units using the same conversion factor $\alpha_0$ for all six intervals. The readout model accounts for the dependence of the physical conversion factor on the resonance frequency in each interval (SM Sec.~\ref{sec:sm_calibration_dependencies}).

For uncertainty propagation, an input condition is one set of parameter values drawn from the assumed measurement and calibration uncertainty distributions. We use 64 such conditions for the spectra and covariance comparisons. Within each verification interval, we use a local model: its resonance frequency and damping are constant.

\begin{figure*}[t]
\centering
\includegraphics[width=0.88\textwidth,height=0.68\textheight,keepaspectratio]{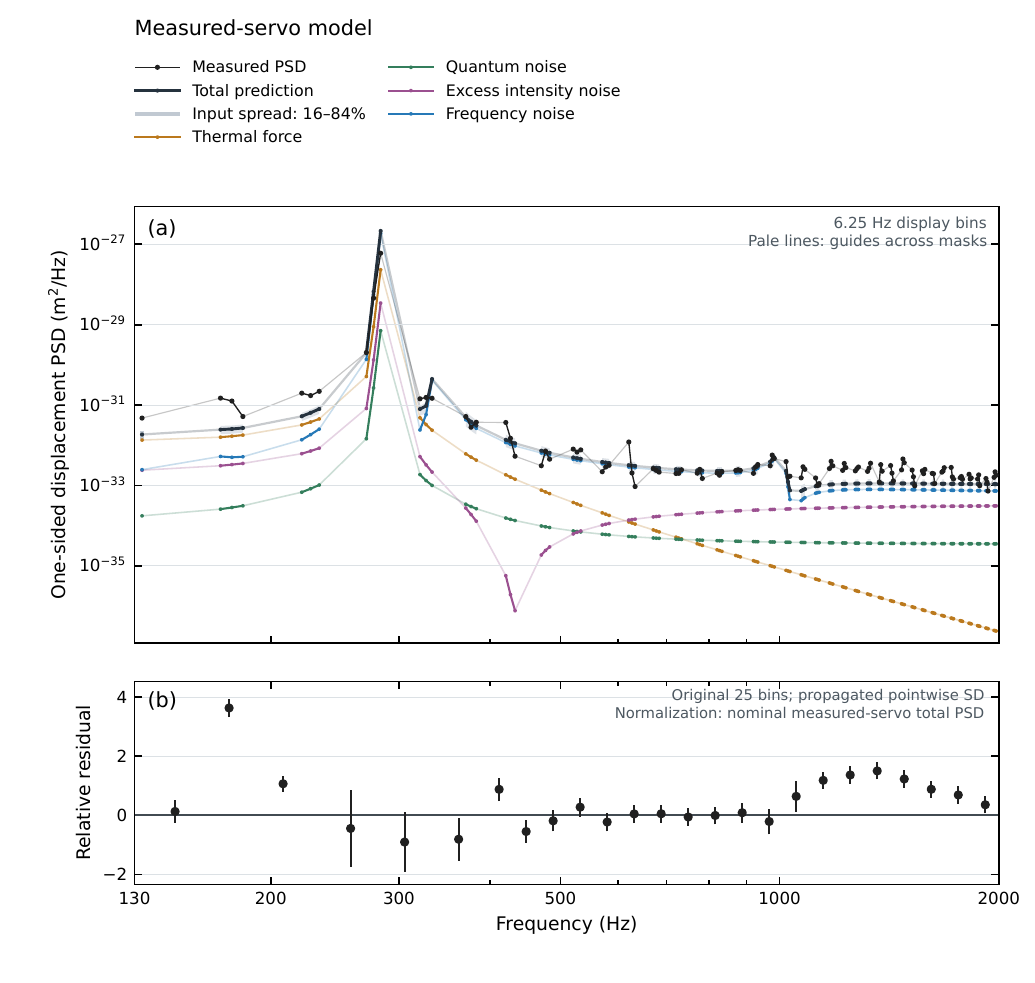}
\caption{Noise budget and residual for the six verification intervals. (a) Measured displacement-referred detector PSD and the thermal, quantum, intensity, and frequency-noise contributions of the adopted model; each classical optical contribution includes its correlated force and direct-readout paths. Predictions include the finite interval and Hann window and are averaged over the 64 input conditions; shading gives the pointwise 16th--84th percentile range of their sum. (b) Measured-minus-predicted PSD divided by the nominal total prediction, with pointwise one-standard-deviation error bars combining finite-record and propagated input uncertainty. Both panels use the nominal displacement calibration. Pale lines across masked gaps are visual guides only. Panel (a) uses 112 common frequency bins and panel (b) uses 25.}
\label{fig:noise_budget_residual}
\end{figure*}

\subsection{Spectral processing and the comparison target}

\begingroup\widowpenalty=10000

To compare reconstruction and prediction under the same processing, we restrict both to the analysis band $B=[130,2000]\,\mathrm{Hz}$, which includes the optically shifted mechanical resonance.
A broad unexplained feature appears near $90\,\mathrm{Hz}$ below this
range; the available model constraints also limit a reliable extension above approximately $2\,\mathrm{kHz}$. This common band is used for both prescriptions.
Both prescriptions are applied to the same record, so the uncertainty of their difference includes the correlation between their reconstructed statistics. Within this band, the largest positive spectral residuals occur around $170$--$240\,\mathrm{Hz}$ and $1100$--$1800\,\mathrm{Hz}$ [Figs.~\ref{fig:interval_psd} and \ref{fig:noise_budget_residual}(b)]. In the model, thermal force dominates the output spectrum below about $250\,\mathrm{Hz}$ and frequency noise above about $330\,\mathrm{Hz}$ [Fig.~\ref{fig:noise_budget_residual}(a)].

Each comparison involves three quantities. The first is the full-band value: $V_2$ for the conventional prescription and $V$ for the three-trajectory prescription. The second is its finite-band prediction, $P_2$ or $P_3$: the integral over $B$ of the spectra of that prescription [Eqs.~\eqref{eq:three_spectrum} and \eqref{eq:theory_band_integral}], calculated for a stationary model with filters derived from that model. For this band, $P_3$ abbreviates $P_{3,B}$.

The third is the experimental statistic, $E_2$ or $E_3$: the corresponding finite-record spectral integral with the continuum correction described below, using filters whose gains are held at their stationary model values within each interval. When $\bm b_c$ is used, the same procedure gives $E_{3c}$ and $P_{3c}=P_{3c,B}$.

Figure~\ref{fig:apparatus_estimator}(b) summarizes the processing, which is identical for experimental and simulated records. Narrow peaks at harmonics of the approximately $50$-Hz mains frequency are treated by notch filtering. We evaluate spectra in the central $80\%$ of each interval to reduce endpoint effects, giving $1.64$ s in total, and average the six interval estimates with equal weights. SM Sec.~\ref{sec:sm_implementation} gives the preprocessing passband, filter discretization, and initialization. The smoothing-error covariance deviated from its stationary value by at most $0.333\%$ at the interval midpoints and by as much as $16.9\%$ at the latest times retained for analysis (SM Sec.~\ref{sec:sm_future_information}). We therefore estimated the finite-record mean shift and uncertainty using simulated records with the same filter initialization and endpoint trimming as the measured records.\par\endgroup

Notch filtering attenuates the continuum (the broadband spectrum apart from the mains lines) and leaves residual lines. We compensate for the attenuation and subtract the residual lines (SM Sec.~\ref{sec:sm_line_processing}); $E_2$, $E_3$, and $E_{3c}$ denote the resulting continuum-corrected statistics. Simulations quantify the remaining difference between each statistic and its prediction caused by finite-record processing (Sec.~\ref{sec:results}\,A).

\section{Results}\label{sec:results}

This section compares the measured bias difference and conditional covariance with their model predictions, using the reference comparison defined in Sec.~\ref{sec:results}\,A. We then characterize the reconstructed squeezing and test the predicted changes when the analysis band is varied.

\subsection{Finite-band comparisons and uncertainty}

Each comparison asks one question: is the difference between a measured statistic and its prediction within the scatter expected from the uncertainty of the measured inputs and the finite record length? We answer with one number, the distance/radius ratio. The numerator is the distance of the difference from zero, in units of its expected scatter and including the correlations among its three entries. The denominator is the radius that contains 95\% of the expected fluctuations, obtained by simulation as described below. A ratio at or below one means that the difference lies within this empirical reference region. All five comparisons in Table~\ref{tab:revised_p} meet this condition, and the upper ends of their numerical stability ranges, obtained by resampling the input conditions, remain below one. All five comparisons also informed the preliminary selection of the frequency-noise model (Sec.~\ref{sec:discussion}\,A).

The table contains two kinds of comparison. The two prescriptions are compared using the future-only center $\bm b_0$, on which the conventional prescription is defined. The comparison with $V$ uses the past-conditioned center $\bm b_c$. The standard uncertainties of $\widetilde E_3-P_3$ with $\bm b_0$ are $(1545,70151,39547)$ in component order $(qq,pp,qp)$, exceeding the corresponding entries of $V$. For $\widetilde E_{3c}-P_{3c}$ with $\bm b_c$, they are $(337,236,161)$, about one quarter of the entries of $V$. SM Sec.~\ref{sec:sm_empirical_uncertainty} gives the uncertainty calculation. We use this component order throughout.

The expected scatter is obtained by simulation. We draw 64 input conditions from the uncertainty distributions of the measured inputs (SM Secs.~\ref{sec:sm_calibration_dependencies} and \ref{sec:sm_noise_scope}). Because the filters are derived from the model, each condition changes both the reconstruction and its prediction, so we reprocess the experimental record and recompute the predictions for each condition. We express all matrices in the common reference quadratures before averaging over the conditions. For each input condition, we simulate 64 complete records and process them in the same way as the experimental record, for a total of 4096 simulated records. We use 32 records per condition (the pilot records) to estimate the mean difference between each finite-record statistic and its corresponding finite-band prediction. We subtract this mean difference, the pilot mean correction, from the experimental statistic and denote the result with a tilde, as in $\widetilde E_{3c}$. The other 32 records per condition (the evaluation records) provide an independent estimate of the record-to-record scatter (SM Sec.~\ref{sec:sm_empirical_uncertainty}).

Averaged over the 64 input conditions, the pilot correction subtracted from $E_{3c}$ is approximately $(120,40,10)$ in $(qq,pp,qp)$ order, or $10\%$, $6\%$, and $2\%$ of the corresponding entries of $P_{3c}$. The shortest of the six verification intervals makes the largest contribution to each entry.

Let $\bm r$ be the observed difference vector before the correction and $\bm\mu_0$ the correction, each averaged over the input conditions. Let $C_0$ combine the variation between input conditions with the finite-record covariance within each condition. The squared distance is
\begin{equation}
 \mathcal T(\bm r)=(\bm r-\bm\mu_0)^{\mathsf T}
                  C_0^{-1}(\bm r-\bm\mu_0).
 \label{eq:main_residual_statistic}
\end{equation}
The centered reference vectors, which represent the expected fluctuations, combine input variation and finite-record fluctuations, each with its mean removed. The squared reference radius is the 95th percentile of their squared distances.

\begin{table}[t]

\caption{Correlated residual distance divided by its empirical 95\% reference radius. Values at or below one lie within the reference region. All residuals include the pilot mean correction; the $\widetilde E_{3c}-V$ comparison uses the same correction as $\widetilde E_{3c}-P_{3c}$. The final column gives the 95\% range obtained by resampling entire input-condition groups, which assesses numerical stability of the Monte Carlo estimate.}
\label{tab:revised_p}
\begin{ruledtabular}
\begin{tabular}{lcc}
Comparison & Distance/radius & Numerical range \\
$\widetilde E_2-P_2$ & $0.49$ & $[0.41,0.62]$ \\
$\widetilde E_3-P_3$ & $0.79$ & $[0.63,0.93]$ \\
$(\widetilde E_2-\widetilde E_3)-(P_2-P_3)$ & $0.57$ & $[0.44,0.74]$ \\
$\widetilde E_{3c}-P_{3c}$ & $0.69$ & $[0.62,0.86]$ \\
$\widetilde E_{3c}-V$ & $0.67$ & $[0.58,0.88]$ \\
\end{tabular}
\end{ruledtabular}

\end{table}

\subsection{Bias difference between the prescriptions}

Applying both prescriptions to the same record cancels the unknown covariance in $\widetilde E_2-\widetilde E_3$; its expectation is the difference between their reconstruction biases.

In the position variance, the conventional result exceeds the three-trajectory result by $(5.3\pm3.6)\times10^{5}$, against a predicted $4.0\times10^{5}$. For scale, $V_{qq}\approx1.3\times10^{3}$: the two prescriptions differ by about 400 times the quantity they are meant to estimate. The three entries of the difference are strongly correlated, so we compare them jointly. The joint distance/radius ratio is $1.44$ relative to zero and $0.57$ relative to the predicted difference $P_2-P_3$, with each comparison using its own propagated covariance and empirical reference radius. The measured difference lies outside the 95\% reference region about zero and within the region about the predicted difference (Table~\ref{tab:revised_p}). The reproducibility package gives the three entries of the difference and the numerical stability ranges for both ratios.

This test concerns the finite-band statistics; the analytic bias of Sec.~\ref{sec:colored_bias_mechanism} refers to the full band.

\subsection{Conditional covariance and finite bandwidth}
\label{sec:experimental_band_examples}

\textit{Comparison with $V$.} The past-conditioned reconstruction agrees with the conditional covariance $V$ within its standard uncertainty in all three entries (Table~\ref{tab:revised_covariance}); the distance/radius of $\widetilde E_{3c}-V$ is $0.67$, inside the 95\% reference region (Table~\ref{tab:revised_p}).

The analysis band also enters this comparison. Adding and subtracting $P_{3c}$ gives
\begin{equation}
 \widetilde E_{3c}-V=(\widetilde E_{3c}-P_{3c})+(P_{3c}-V).
 \label{eq:experimental_covariance_decomposition}
\end{equation}
The first term is the departure of the reconstruction from its finite-band prediction. The second is the finite-band offset, the amount by which the limited band shifts the expected reconstruction away from $V$; its relative size is $\epsilon_{B,c}$ [Eq.~\eqref{eq:band_accuracy} with $\bm b_c$]. The mean offset is approximately $(-160,-400,-120)$ in component order $(qq,pp,qp)$: in this band, all three entries are expected to be smaller than those of $V$. For each input condition, $\epsilon_{B,c}$ is the maximum absolute relative variance difference over quadrature directions. Its mean across the 64 conditions is $44\%$ (range $38$--$60\%$). This offset is comparable to the standard uncertainty of $\widetilde E_{3c}$, $(350,230,150)$. Accordingly, the observed reconstruction is consistent with both $P_{3c}$ and $V$ in the comparisons of Table~\ref{tab:revised_p}. The bandwidth scan below separates the two: for upper limits of $200$ and $300\,\mathrm{Hz}$, the reconstruction follows the finite-band prediction and lies $1.9$--$5.1$ standard uncertainties below $V$.

\begin{figure}[t]

\centering
\includegraphics[width=\columnwidth]{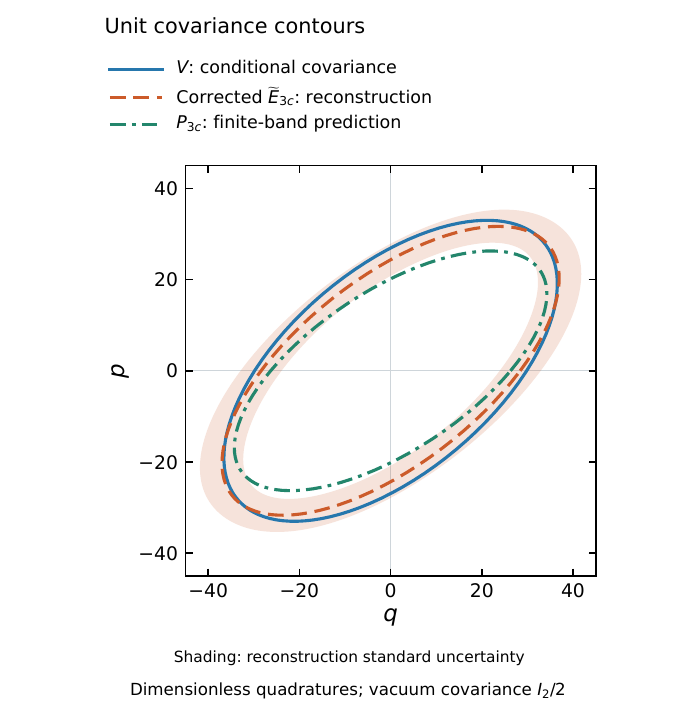}
\caption{Covariance contours in the common dimensionless position and momentum quadratures, with vacuum covariance $I_2/2$: mean conditional covariance $V$ (blue solid), corrected experimental reconstruction $\widetilde E_{3c}$ (orange dashed), and finite-band prediction $P_{3c}$ (green dash-dotted), with semiaxes equal to the square roots of the eigenvalues of the mean matrices underlying Table~\ref{tab:revised_covariance}. The shaded band shows the pointwise standard uncertainty of the reconstructed contour.}
\label{fig:revised_covariance}

\end{figure}

\textit{Squeezing.} The reconstructed state is thermally squeezed (Fig.~\ref{fig:revised_covariance}): its uncertainty is smaller along one quadrature direction than along the orthogonal one. Its principal variances, the smallest and largest variances over all quadrature directions, are approximately $(410,1960)$, a ratio of $4.7\pm0.7$; those of $V$ are $(500,1920)$, a ratio of $3.8$. A ratio of one would mean no squeezing. The squeezing is determined mainly by the cross covariance. This entry is $749\pm147$, about five standard uncertainties from zero, whereas the diagonal entries differ by $364\pm257$, about $1.4$ standard uncertainties. The diagonal entries alone would give a variance ratio of only $1.36\pm0.27$. The purity under the Gaussian-state assumption is $1/(2\sqrt{\det\widetilde E_{3c}})=(5.6\pm1.4)\times10^{-4}$, where a pure state has purity one. This purity corresponds to an equivalent thermal occupation of about $900$, obtained after mathematically removing the squeezing. It is about seven orders of magnitude below the equilibrium occupation of $2.2\times10^{10}$ at $300\,\mathrm K$ and the reference frequency $283.5\,\mathrm{Hz}$. The uncertainties of the ratio and the purity are propagated to first order from the joint uncertainty of the three entries (SM Sec.~\ref{sec:sm_empirical_uncertainty}).

\begin{table}[t]

\caption{Conditional covariance comparison in the common dimensionless reference quadratures. Entries are means over the 64 input conditions, rounded to the nearest ten. The reproducibility package gives the unrounded means. ``Before'' and ``after'' refer only to the pilot mean correction; the continuum correction of Sec.~\ref{sec:experiment}\,C has already been applied. Uncertainties are standard uncertainties that include input and finite-record variation; the last row propagates them jointly.}
\label{tab:revised_covariance}
\begin{ruledtabular}
\begin{tabular}{lrrr}
Quantity & $qq$ & $pp$ & $qp$ \\
$E_{3c}$ before mean correction & $1490$ & $1040$ & $760$ \\
$\widetilde E_{3c}$ after mean correction & $1370$ & $1000$ & $750$ \\
Standard uncertainty of $\widetilde E_{3c}$ & $350$ & $230$ & $150$ \\
$P_{3c}$ & $1170$ & $690$ & $580$ \\
$V$ & $1340$ & $1090$ & $700$ \\
Corrected $\widetilde E_{3c}-V$ & $30\pm360$ & $-90\pm270$ & $50\pm180$ \\
\end{tabular}
\end{ruledtabular}

\end{table}

\begin{figure*}[t]
\centering
\includegraphics[width=\textwidth]{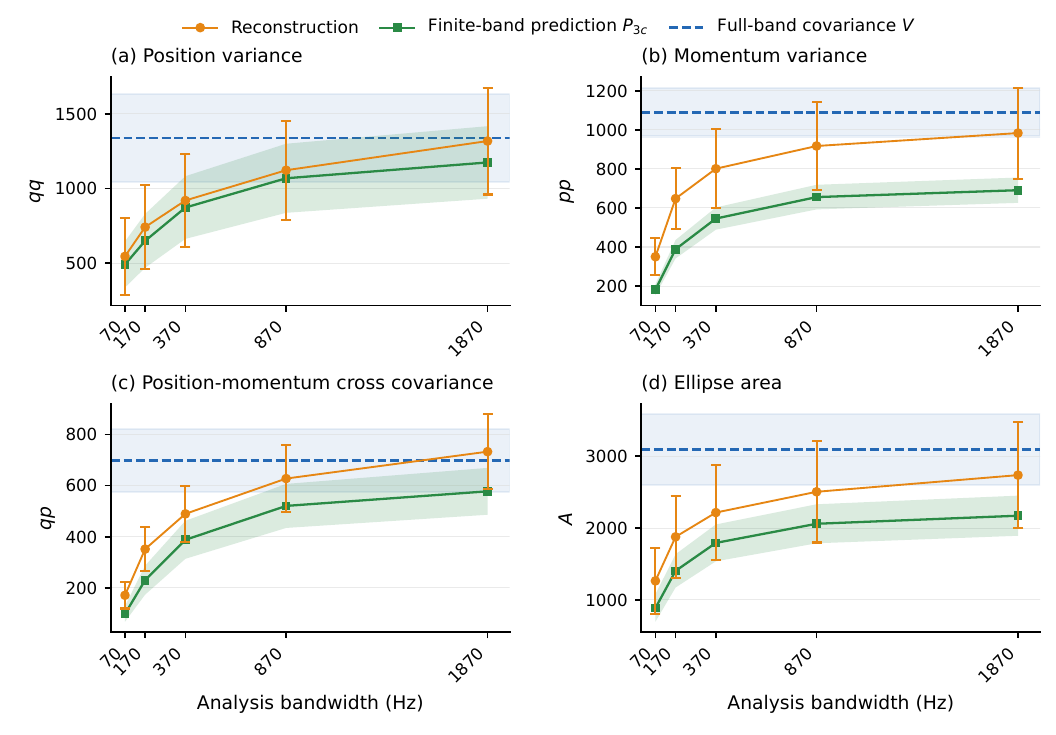}
\caption{Bandwidth dependence of the reconstructed covariance. The lower frequency limit is fixed at $130\,\mathrm{Hz}$; the upper limits are $200$, $300$, $500$, $1000$, and $2000\,\mathrm{Hz}$. The horizontal axis gives the corresponding bandwidths, $70$--$1870\,\mathrm{Hz}$. Circles show the reconstruction, squares joined by lines the finite-band predictions, and dashed horizontal lines the common full-band covariance $V$. Error bars and shading show pointwise standard uncertainties. The scan uses analytic finite-record mean corrections, whereas Fig.~\ref{fig:revised_covariance} and Table~\ref{tab:revised_covariance} use Monte Carlo pilot corrections; at $130$--$2000\,\mathrm{Hz}$ the two agree within $0.15$ standard uncertainties. Panels (a)--(c) show the $qq$, $pp$, and $qp$ entries. Panel (d) shows the ellipse area, labeled $A$ in the panel and calculated as $\pi$ times the square root of the determinant of each mean matrix, with uncertainty propagated from all three entries. The comparisons of bandwidth changes include correlations between bands (SM Sec.~\ref{sec:sm_bandwidth_scan}).}
\label{fig:bandwidth_scan}
\end{figure*}

\textit{Bandwidth scan.} We reanalyze the same record with the lower band edge fixed at $130\,\mathrm{Hz}$ and upper edges of $200$, $300$, $500$, $1000$, and $2000\,\mathrm{Hz}$ (Fig.~\ref{fig:bandwidth_scan}). We use the same model and full-band covariance $V$ throughout. The reconstruction is precise enough to resolve the effect of bandwidth on all three covariance entries. When we widen the analysis band from $130$--$200\,\mathrm{Hz}$ to $130$--$2000\,\mathrm{Hz}$, the measured changes range from $3.9$ to $5.2$ times their respective standard uncertainties. The reconstructed ellipse area increases by $1473\pm365$ [Fig.~\ref{fig:bandwidth_scan}(d)]. The predicted increase is $1286\pm140$. After accounting for correlations, the difference between the measured and predicted increases is $187\pm386$, consistent with zero.

At each upper band edge, all three reconstructed entries lie $0.2$--$1.8$ standard uncertainties above their finite-band predictions. These residuals are strongly correlated across the nested bands because the bands share data (SM Sec.~\ref{sec:sm_bandwidth_scan}). As the upper edge is raised in steps to $1000\,\mathrm{Hz}$, the measured change at each step agrees with the predicted change within $1.3$ standard uncertainties. Adding $1000$--$2000\,\mathrm{Hz}$, however, increases each entry by about two standard uncertainties more than predicted. These three residuals are strongly correlated, and the measured PSD also exceeds the model over most of this frequency range [Fig.~\ref{fig:noise_budget_residual}(b)].

The two narrowest bands distinguish the finite-band prediction from $V$. For upper edges of $200$ and $300\,\mathrm{Hz}$, the reconstructed entries lie $1.9$--$5.1$ standard uncertainties below the corresponding entries of $V$. At $500\,\mathrm{Hz}$ they lie $1.1$--$1.3$ standard uncertainties below $V$, and at $1000$ and $2000\,\mathrm{Hz}$ they are within about $0.6$ standard uncertainties of $V$. Thus, for the three widest bands, the reconstructed entries are consistent with both $V$ and the finite-band prediction. The comparisons with $V$ account for the uncertainty in the measured inputs shared by the reconstruction and $V$.

\section{Discussion}\label{sec:discussion}

\subsection{Scope and limitations}

In Table~\ref{tab:revised_p}, the adopted model is consistent with the finite-band statistics of both reconstruction methods and with their difference on the same record. These comparisons test weighted integrals of the output PSD (SM Sec.~\ref{sec:sm_diagnostics}). Figure~\ref{fig:noise_budget_residual}(b) shows the residuals as a function of frequency. The origin of the excess observed between $1000$ and $2000\,\mathrm{Hz}$ remains unknown (Sec.~\ref{sec:experimental_band_examples}).

Two analysis choices were made with knowledge of the data. We selected the laser frequency-noise model in a preliminary screening constrained by the measured servo response and output spectrum. The screening used the same five covariance comparisons later reported in Table~\ref{tab:revised_p}. After data collection, we selected six intervals in which the optical-spring resonance was comparatively stable. We estimated the uncertainties by varying the calibration and other measured inputs and simulating repeated records, while keeping the noise-model structure and the six time windows fixed. Measurements of the suspension, laser intensity noise before the cavity, and the servo response constrain the model. Different combinations of internal noise sources can nevertheless produce the same output spectrum, and a simultaneous out-of-loop measurement of the stabilized laser frequency was unavailable.

Frequency noise can drive the mirror while partly canceling in the readout, so part of the motion is only weakly visible in the record. A reconstruction that discarded weakly determined directions would then underestimate the covariance. Here, after conditioning on the past, the future record reduces the uncertainty in every direction of mechanical phase space, and the reconstruction retains all directions. The full-band identity is satisfied to numerical precision with stationary model moments (SM Sec.~\ref{sec:sm_future_information_diagnostics}).

With the lower band edge fixed at $130\,\mathrm{Hz}$, the two narrowest bands resolve the effect of finite bandwidth. For upper edges of $200$ and $300\,\mathrm{Hz}$, all three reconstructed covariance entries agree with their finite-band predictions and lie below the corresponding entries of the full-band conditional covariance $V$. Although the measurements no longer resolve the finite-band offset from $V$ at upper band edges of $500$, $1000$, and $2000\,\mathrm{Hz}$, all three reconstructed covariance entries are consistent with both the finite-band prediction and $V$ within the experimental uncertainties. At $130$--$2000\,\mathrm{Hz}$, the same reconstruction is also consistent with both $P_{3c}$ and $V$ when the three entries are compared jointly (Table~\ref{tab:revised_p}). Finally, omitting only the pilot mean correction reduces the principal-variance ratio of the mean reconstructed covariance from $4.72$ to $4.33$ (Table~\ref{tab:revised_covariance}). Thermal squeezing is therefore evident even before the correction, although its estimated magnitude changes.

\subsection{Outlook for quantum-state verification}
\label{sec:quantum_outlook}

The three-trajectory reconstruction is directly applicable to verifying the states prepared in future experiments aimed at generating entanglement between two mirrors~\cite{MuellerEbhardt2008,Miki2023}. Sawada \textit{et al.}~\cite{Sawada2026} measured a free-pendulum energy-decay rate of $\Gamma/(2\pi)=361(39)\,\mathrm{nHz}$ for a suspension with a stepped fiber diameter, about $1/130$ of the rate of the suspension used here. Their forecast applies the past-conditioned reconstruction [Eqs.~\eqref{eq:main_conditional_center} and \eqref{eq:conditional_band_prediction}] to a multimode model. For the stepped suspension at the selected optical settings, every covariance within the forecast's error bounds that satisfies the quantum uncertainty relation corresponds to an entangled Gaussian state.

The agreement between our $130$--$2000\,\mathrm{Hz}$ reconstruction and the model's full-band conditional covariance $V$ within the propagated uncertainties (Tables~\ref{tab:revised_p} and \ref{tab:revised_covariance}) motivates the $130$--$1990\,\mathrm{Hz}$ verification band used in the two-mirror forecast of Ref.~\cite{Sawada2026}. The forecast uses our $10.5\%$ relative standard uncertainty in displacement scale and treats the $1.64\,\mathrm{s}$ retained across our six intervals as one continuous stationary record. For the stepped suspension at the selected optical settings, our calibration uncertainty is below the $19.4\%$ limit for entanglement verification with all other inputs fixed.

\section{Conclusions}\label{sec:conclusions}

The three-trajectory reconstruction separates the conditional covariance from the future-likelihood width. One rule gives the position variance, the momentum variance, and their cross covariance, allowing the magnitude and orientation of squeezing to be evaluated together. An analytic oscillator model shows how uncertainty in the initial amplitudes of the colored force contributes to the conventional overestimate of the position variance. For a fixed stationary model, conditioning the future likelihood on the past gives an upper bound on the error from omitted frequencies. This bound does not increase as the analysis band widens.

Applied to the center-of-mass motion of a $7.71$-mg mirror, the reconstruction yields a thermally squeezed conditional state with a principal-variance ratio of $4.7\pm0.7$. The inferred purity is $(5.6\pm1.4)\times10^{-4}$, corresponding to an equivalent thermal occupation of about 900 phonons after mathematically removing the squeezing. Both the reconstructed covariance matrix and the measured bias difference between the two prescriptions agree with the adopted model within the propagated uncertainties. The bandwidth scan resolves changes in all three covariance entries. As the upper band edge is raised in steps to $1\,\mathrm{kHz}$, the measured change at each step agrees with the finite-band prediction.

These results show how a finite-band measurement, compared with the prediction for the same band, can characterize a conditional mechanical state. They provide a basis for future verification of quantum squeezing and entanglement with improved noise control and calibration.

\begin{acknowledgments}
We thank Guanhao Huang and Naoki Yamamoto for valuable comments.
K.S. is supported by the RIKEN TRIP initiative (RIKEN Quantum).
N.M. is supported by JST FOREST Grant No.~JPMJFR202X.
\par
We used GPT-6 Astra (OpenAI) and Claude Fable 5.1 and 5.2 (Anthropic) to assist with analysis and manuscript editing. The authors verified the results, reviewed the AI-assisted text, and take responsibility for the final content.
\par
N.M. conceived the three-trajectory estimator, reanalyzed the experimental data, and prepared the manuscript. K.S. and N.M. implemented the method and performed the numerical analysis. Both authors discussed the results and approved the final manuscript.
\end{acknowledgments}
\section*{Data and code availability}

The original detector-voltage record and earlier analysis files are available on Zenodo~\cite{SakaiMatsumoto2026Data}. The reproducibility package accompanying this revision~\cite{MatsumotoUnbiasedReproducibility} contains the adopted model, calibration inputs, reconstruction code, uncertainty calculations, and figure data.

\appendix
\section{Augmented likelihood and covariance identities}
\label{app:identity}

This appendix proves the covariance identity in the augmented state, then gives the mechanical projection, the conditions for a likelihood center to exist, and the spectral form used in reconstruction.

\subsection{Augmented likelihood and Gaussian orthogonality}
\label{sec:identity_proof}

To prove the identity, we express the future record as an observation of the augmented state and separate its noise from the forward estimation error. All covariances in this appendix describe Gaussian estimation errors or classical reconstructed trajectories; the mechanical quantum covariance uses symmetrically ordered moments as in Eq.~\eqref{eq:quantum_conditional_covariance}. We use the state and record partition of Sec.~\ref{sec:theory}. The forward error covariance $P$ is positive definite, and we first assume that the augmented future information matrix $J$ is also positive definite.

Collect the future observations into a vector $\bm y_+$. The augmented Markov model gives
\begin{equation}
 \bm y_+=\mathcal H\bm z_t+\bm\epsilon,
 \qquad \operatorname{Cov}(\bm\epsilon)=\Sigma_\epsilon,
 \label{eq:app_future_observation}
\end{equation}
where $\mathcal H$ maps the state $\bm z_t$ to the conditional mean of the future record given $\bm z_t$. Because the OU coordinates belong to $\bm z_t$, $\bm\epsilon$ contains only subsequent process and readout increments, whose correlations enter $\Sigma_\epsilon$; it is therefore independent of $\bm z_t$ and the past record. The Gaussian information parameters are
\begin{equation}
 J=\mathcal H^{\mathsf T}\Sigma_\epsilon^{-1}\mathcal H,
 \qquad \bm h=\mathcal H^{\mathsf T}\Sigma_\epsilon^{-1}\bm y_+.
 \label{eq:app_future_information}
\end{equation}
Here $\Sigma_\epsilon$ is the future-record noise covariance, positive definite on the retained observation space.
Substituting Eq.~\eqref{eq:app_future_observation} into Eq.~\eqref{eq:app_future_information} gives the augmented likelihood center
\begin{equation}
 \bm b_z=J^{-1}\bm h
 =\bm z_t+J^{-1}\mathcal H^{\mathsf T}\Sigma_\epsilon^{-1}\bm\epsilon.
 \label{eq:app_augmented_center}
\end{equation}
Its error $\bm e_b=\bm z_t-\bm b_z$ has covariance $J^{-1}$ and zero cross covariance with the forward error $\bm e_f=\bm z_t-\bm f_z$.
To see both statements, substitute Eq.~\eqref{eq:app_augmented_center}:
the state cancels from $\bm e_b$, leaving a linear function of future
noise only. Using the symmetry of $J$ and $\Sigma_\epsilon$ gives
\begin{align}
 \operatorname{Cov}(\bm e_b)
 &=J^{-1}\mathcal H^{\mathsf T}\Sigma_\epsilon^{-1}
   \Sigma_\epsilon\Sigma_\epsilon^{-1}\mathcal HJ^{-1}\notag\\
 &=J^{-1}JJ^{-1}=J^{-1},\qquad
 \operatorname{Cov}(\bm e_f,\bm e_b)=0.
 \label{eq:app_center_error_covariance}
\end{align}
The zero cross covariance follows from the independence of the future increments from the state and past record. Since $\bm f_z-\bm b_z=\bm e_b-\bm e_f$,
the two error widths therefore add. Consequently,
\begin{equation}
 \operatorname{Cov}(\bm f_z-\bm b_z)=P+J^{-1}.
 \label{eq:app_forward_backward}
\end{equation}

Let $\bm e_s=\bm z_t-\bm s_z$ be the smoothing error, with covariance $P_s$. Because $\mathbb E[\bm e_s\mid\mathcal Y_-,\mathcal Y_+]=0$, the smoothing error is uncorrelated with every square-integrable function of the full record, in particular with $\bm s_z-\bm f_z$ and $\bm s_z-\bm b_z$. The decompositions
\begin{align}
 \bm e_f&=\bm e_s+(\bm s_z-\bm f_z),\notag\\
 \bm e_b&=\bm e_s+(\bm s_z-\bm b_z)
\end{align}
therefore give the cross covariances needed below:
\begin{equation}
 \operatorname{Cov}(\bm e_s,\bm e_f)
 =\operatorname{Cov}(\bm e_s,\bm e_b)=P_s.
 \label{eq:app_smoother_error_cross}
\end{equation}
Taking the covariance of each decomposition now gives
\begin{align}
 \operatorname{Cov}(\bm s_z-\bm f_z)&=P-P_s,\notag\\
 \operatorname{Cov}(\bm s_z-\bm b_z)&=J^{-1}-P_s.
 \label{eq:app_smoothing_differences}
\end{align}
Combining Eqs.~\eqref{eq:app_forward_backward} and \eqref{eq:app_smoothing_differences} gives the augmented identity
\begin{align}
 P=\tfrac12\{&\operatorname{Cov}(\bm f_z-\bm b_z)
 +\operatorname{Cov}(\bm s_z-\bm f_z)\notag\\
 &-\operatorname{Cov}(\bm s_z-\bm b_z)\}.
 \label{eq:app_identity_proof}
\end{align}
There is also the nonsubtractive form
\begin{equation}
 \operatorname{Cov}(\bm f_z-\bm s_z,\bm f_z-\bm b_z)=P.
 \label{eq:app_augmented_cross}
\end{equation}
Applying $\Pi$ and $\Pi^{\mathsf T}$ proves Eqs.~\eqref{eq:difference_covariance_values} and \eqref{eq:three_identity}, and gives the mechanical cross-covariance form
\begin{equation}
 V=\operatorname{sym}\operatorname{Cov}
       (\bm f-\bm s,\bm f-\bm b_0),
 \label{eq:cross_identity}
\end{equation}
where $\operatorname{sym}X=(X+X^{\mathsf T})/2$. Equation~\eqref{eq:cross_identity} is algebraically equivalent to Eq.~\eqref{eq:three_identity} and is used in the numerical implementation.

In data assimilation~\cite{Desroziers2005,Todling2022}, $\bm f_z$ is the background (prediction before the observation), $\bm b_z$ the observation, and $\bm s_z$ the analysis (updated estimate). Here the future likelihood supplies the observation, with error independent of the forward prediction error.

\subsection{Projection to the mechanical coordinates}
\label{sec:mechanical_projection}

\begingroup\displaywidowpenalty=10000
To obtain the mechanical future likelihood, we integrate out the auxiliary variables with a flat measure. Its center is $\bm b_0=\Pi\bm b_z$. Partition $J$ and $\bm h$ according to $(\bm x,\bm\eta)$. For nonsingular $J_{\eta\eta}$, this integration gives the Schur-complement precision $J_0$ and center:
\begin{align}
 J_0&=J_{xx}-J_{x\eta}J_{\eta\eta}^{-1}J_{\eta x},\notag\\
 \bm h_0&=\bm h_x-J_{x\eta}J_{\eta\eta}^{-1}\bm h_\eta,\notag\\
 \bm b_0&=J_0^{-1}\bm h_0,\qquad V^{(\ell)}=J_0^{-1}.
 \label{eq:app_projected_information}
\end{align}
The square completion behind this elimination is
\begin{align}
 \bar{\bm\eta}(\bm x)
 &=J_{\eta\eta}^{-1}(\bm h_\eta-J_{\eta x}\bm x),\notag\\
 \log p(\mathcal Y_+\mid\bm x,\bm\eta)
 &=-\tfrac12[\bm\eta-\bar{\bm\eta}(\bm x)]^{\mathsf T}
 J_{\eta\eta}[\bm\eta-\bar{\bm\eta}(\bm x)]\notag\\
 &\quad-\tfrac12\bm x^{\mathsf T}J_0\bm x
 +\bm h_0^{\mathsf T}\bm x+\text{constant}.
 \label{eq:app_likelihood_completion}
\end{align}
Integrating the Gaussian factor in $\bm\eta$ with a flat measure gives an
$\bm x$-independent factor, leaving precision $J_0$ and linear coefficient
$\bm h_0$. The mechanical center is therefore $\Pi J^{-1}\bm h$ and depends on the future record alone.\par\endgroup

The smoother must still be constructed in the augmented state using Eq.~\eqref{eq:augmented_smoother}. In general,
$U\ne(V^{-1}+[V^{(\ell)}]^{-1})^{-1}$: the mechanical forward and future factors obtained by separate elimination of correlated auxiliary variables cannot simply be multiplied as independent two-dimensional factors. Retaining the mechanical--auxiliary covariance block $P_{x\eta}$ of $P$ and the augmented information before projection accounts for those correlations.

\subsection{Rank and record endpoint}
\label{sec:identity_rank_endpoint}

We now allow $J$ to be singular. The future record determines a mechanical likelihood center precisely when a matrix $B_0$ satisfies
\begin{equation}
 B_0J=\Pi,
 \label{eq:app_target_rank}
\end{equation}
that is, when no unobservable augmented direction has a mechanical projection. Then $\bm b_0=B_0\bm h$ has likelihood error width $B_0JB_0^{\mathsf T}$, and the covariance proof applies unchanged. It also applies to any finite collection of target coordinates $\bm x=\Pi\bm z_t$, including several mechanical modes, with $I_2$ replaced by the identity on that target space.

At $T=t$, there are no future observations: $J=0$, $\bm h=0$, and $P_s=P$. At fixed $t$, as $T$ is extended, $J$ is nondecreasing and $P_s$ is nonincreasing in the positive-semidefinite order, while the forward covariance $P$ remains fixed. In every direction admitting a center, the endpoint dependence of the likelihood width cancels between Eqs.~\eqref{eq:app_forward_backward} and \eqref{eq:app_smoothing_differences}. For the stable stationary models considered here, $J$ approaches a limiting future information matrix as $T$ increases; the rate of approach and the directions in which information is available are set by the dynamics and observation. The experimental analysis applies stationary gains to finite intervals and assesses the resulting endpoint effects by complete-record simulations.

\subsection{Mechanical likelihood conditioned on the past}
\label{sec:sm_alternative_definition}

We now derive the past-conditioned center $\bm b_c$ introduced in Eq.~\eqref{eq:main_conditional_center}. The future likelihood is obtained by averaging over the auxiliary variables using their distribution conditioned on the mechanical state and the past record. The use of $\bm b_c$ in choosing the analysis band is illustrated in Appendix~\ref{app:bandwidth}; its experimental reconstruction is used for the conditional-covariance comparison in Sec.~\ref{sec:experimental_band_examples}.

Partition the forward covariance according to the mechanical and auxiliary variables:
\begin{equation}
 P=\begin{pmatrix}V&P_{x\eta}\\
 P_{\eta x}&P_{\eta\eta}\end{pmatrix}.
 \label{eq:app_partition}
\end{equation}
Let $\bm f_\eta$ be the auxiliary part of the forward mean $\bm f_z$. The Gaussian conditional distribution is
\begin{align}
 \bm\eta\mid\bm x,\mathcal Y_- &\sim
 \mathcal N\!\left(\bm f_\eta+K_\eta(\bm x-\bm f),W_\eta\right),\notag\\
 K_\eta&=P_{\eta x}V^{-1},\notag\\
 W_\eta&=P_{\eta\eta}-P_{\eta x}V^{-1}P_{x\eta}.
 \label{eq:app_ou_conditional}
\end{align}
The mechanical future likelihood conditioned on the past is
\begin{align}
 p(\mathcal Y_+\mid\bm x,\mathcal Y_-)
 &=\int p(\mathcal Y_+\mid\bm x,\bm\eta)
 p(\bm\eta\mid\bm x,\mathcal Y_-)\,d\bm\eta\notag\\
 &\propto\frac{p(\bm x\mid\mathcal Y_-,\mathcal Y_+)}
 {p(\bm x\mid\mathcal Y_-)}.
 \label{eq:app_marginalize}
\end{align}
The first expression follows from the augmented Markov property; the second follows from Bayes' rule. The integration retains the residual auxiliary covariance $W_\eta$ from Eq.~\eqref{eq:app_ou_conditional}.

The marginal forward and smoothed densities have means $\bm f,\bm s$ and covariances $V,U$. Their ratio is
\begin{align}
 p(\mathcal Y_+\mid\bm x,\mathcal Y_-)
 &\propto \exp[-\tfrac12\bm x^{\mathsf T}J_c\bm x+
 \bm h_c^{\mathsf T}\bm x],\notag\\
 J_c&=U^{-1}-V^{-1},\qquad
 \bm h_c=U^{-1}\bm s-V^{-1}\bm f.
 \label{eq:conditional_future}
\end{align}
Writing $D=V-U$, a unique center exists when $D$ is positive definite:
\begin{align}
 \bm b_c&=J_c^{-1}\bm h_c
 =\bm f+M(\bm s-\bm f),\notag\\
 M&=VD^{-1},\qquad
 V_c^{(\ell)}=J_c^{-1}=VD^{-1}V-V.
 \label{eq:conditional_center}
\end{align}
Here $V_c^{(\ell)}$ is the conditional-likelihood width. The center $\bm b_c$ uses the past through Eq.~\eqref{eq:app_ou_conditional} and therefore depends on both parts of the record. Moreover, equivalent augmented representations that preserve the joint distribution of the mechanical target and the observation records give the same $\bm f,\bm s,V,U$ and hence the same past-conditioned center $\bm b_c$.

\subsection{Identity, rank, and white-noise limit for the past-conditioned center}
\label{sec:conditional_center_rank}

To verify the identity for the past-conditioned center, we use the forward--smoother difference $\bm a=\bm f-\bm s$. Smoothing orthogonality gives
\begin{equation}
 \operatorname{Cov}(\bm a)=V-U=D.
 \label{eq:app_increment}
\end{equation}
Indeed, the error decomposition $\bm x-\bm f=(\bm x-\bm s)+(\bm s-\bm f)$ has vanishing cross terms because $\bm s-\bm f$ is determined by the full record. Equation~\eqref{eq:conditional_center} implies
$\bm f-\bm b_c=M\bm a$ and $\bm s-\bm b_c=(M-I_2)\bm a$.
The corresponding difference covariances are
\begin{align}
 C_{fb,c}&=MDM^{\mathsf T},\notag\\
 C_{sf,c}&=D,\notag\\
 C_{sb,c}&=(M-I_2)D(M-I_2)^{\mathsf T}.
\end{align}
Since $MD=DM^{\mathsf T}=V$,
\begin{align}
 \tfrac12(C_{fb,c}+C_{sf,c}-C_{sb,c})&=V,\notag\\
 \operatorname{sym}\operatorname{Cov}(\bm a,M\bm a)&=V.
 \label{eq:sm_conditional_identity}
\end{align}
One also obtains $C_{fb,c}=V+V_c^{(\ell)}$ and $C_{sb,c}=V_c^{(\ell)}-U$. The gain $M$ is computed from model covariances; the empirical covariance of $\bm a$ enters only the reconstruction, not $M$. Using that same empirical covariance both in $M=VD^{-1}$ and in the reconstruction would enforce agreement with $V$ algebraically.

The matrices $D$ and $J_c$ have equal rank because $J_c=U^{-1}DV^{-1}$. A unique mechanical center exists only at full rank. Small eigenvalues of $V^{-1/2}DV^{-1/2}$ indicate directions whose variance is only weakly improved by the future record, and warn of an ill-conditioned center.

If the mechanical pair is itself the complete Markov state, $U=(V^{-1}+J)^{-1}$, $J_c=J$, and $\bm b_c=\bm b_0$. For colored noise, both the augmented identity using $\bm b_0$ and Eq.~\eqref{eq:sm_conditional_identity} remain valid, although their frequency weights differ.

\subsection{Stationary spectral form}
\label{sec:spectral_implementation}

To evaluate the identity without subtracting large autospectra, we use a cross spectrum. For stationary filters driven by a scalar record with one-sided PSD $S_y(\nu)$, let $H_a(\nu)$ and $H_d(\nu)$ be the two-component transfer functions producing $\bm f-\bm s$ and $\bm f-\bm b_0$, respectively. With the convention $S_{ad}=H_aS_yH_d^{\dagger}$, the real symmetric reconstruction spectrum is
\begin{equation}
 \operatorname{Re} S_3(\nu)=\operatorname{sym}\operatorname{Re}
 [H_a(\nu)S_y(\nu)H_d^{\dagger}(\nu)].
 \label{eq:app_cross_spectrum}
\end{equation}
Here $\dagger$ denotes conjugate transpose. The opposite cross-spectrum convention changes the imaginary part but leaves this real symmetric quantity unchanged. Equation~\eqref{eq:app_cross_spectrum} equals the real part of Eq.~\eqref{eq:three_spectrum}.

\section{Bias for a finite model of colored force noise}
\label{app:colored_bias}

This appendix derives the bias of the conventional prescription for the
continuous oscillator model of Sec.~\ref{sec:colored_bias_mechanism},
which combines momentum damping at rate $\gamma$, a colored force
represented by a finite number of OU variables, and white direct readout.
The conventional prescription overestimates the position variance by
$R_y(\gamma+\sum_j a_j)$ [Eq.~\eqref{eq:cb_bias}]. Each observable
auxiliary force variable adds $R_y a_j$ to this bias. The same equation
gives the bias in the momentum variance. The off-diagonal component equals
$V_{qp}$ [Eq.~\eqref{eq:cb_conventional_qp}].

This model and the experimental model of Sec.~\ref{sec:experiment}\,A
both represent momentum damping and colored force as separate inputs.
In the experiment, the effective damping rate describes the cooled motion,
and OU variables represent the suspension's structural thermal force.
The experimental model also includes laser intensity and frequency noise,
each of which drives the mirror and enters the readout directly.

To derive the result, we evaluate the forward covariance
(Appendix~\ref{app:colored_bias}\,2) and the future likelihood width
(Appendix~\ref{app:colored_bias}\,3); half their difference gives the
variance biases (Appendix~\ref{app:colored_bias}\,4). All quantities in this
appendix are stationary and full band. The sampled and finite-band
predictions used in the experimental comparison retain the processing
specified in Sec.~\ref{sec:experiment}.

\Needspace{8\baselineskip}
\subsection{Model and information conditions}

\begingroup\displaywidowpenalty=10000
To isolate the colored-force contribution to the bias, we use the finite oscillator model below. Normalize the readout so that its displacement coefficient is one, and
write the augmented state as
$\bm z=(q,p,\eta_1,\ldots,\eta_{N_{\rm OU}})^{\mathsf T}$. The drift and measurement
matrices are
\begin{equation}
 \begin{aligned}
 A&=\begin{pmatrix}
 0&\omega&0^{\mathsf T}\\
 -\omega&-\gamma&\bm c^{\mathsf T}\\
 0&0&-\operatorname{diag}(a_1,\ldots,a_{N_{\rm OU}})
 \end{pmatrix},\\
 C&=\begin{pmatrix}1&0&0^{\mathsf T}\end{pmatrix},
 \end{aligned}
 \label{eq:cb_model}
\end{equation}
where $\omega>0$ is the mechanical angular frequency, $\gamma>0$ is
the momentum damping rate, $a_j>0$ are OU decay rates, and $c_j$ couple
the OU coordinates to the dimensionless momentum equation; in terms of the force coefficients $F_j$ of Eq.~\eqref{eq:exp_ou}, $c_j=F_j/\sqrt{\hbar m\omega_i}$. With each OU coordinate
normalized to unit stationary variance, take
\begin{equation}
 \begin{aligned}
 \mathcal Q&=\operatorname{diag}(0,Q_p,2a_1,\ldots,2a_{N_{\rm OU}}),\\
 \mathcal L&=\varsigma_{py}\bm e_p,\qquad \mathcal R=R_y>0.
 \end{aligned}
 \label{eq:cb_noise}
\end{equation}
Here $Q_p$ is the direct white-force diffusion, $\varsigma_{py}$ is its cross covariance
with the readout increment, and $\bm e_p$ selects momentum. The joint
noise covariance requires $Q_p\ge \varsigma_{py}^2/R_y$. There is no
direct diffusion in $q$. This force-only colored-noise model illustrates the bias mechanism. The experimental model also includes colored noise in the direct readout and is evaluated numerically with the full augmented filters.\par\endgroup

We assume a finite observable state-space representation of the mechanical
motion and colored force, and stabilizing Riccati solutions. For Eq.~\eqref{eq:cb_model}, distinct $a_j$ with nonzero $c_j$
ensure observability of the auxiliary force variables. Uncoupled or redundant
coordinates must be removed before taking the inverse of the future
information matrix. Let $P$ be the augmented forward covariance and
$V_z^{(\ell)}=J^{-1}$ the augmented future likelihood width. Their stationary
equations are
\begin{align}
 K&=(PC^{\mathsf T}+\mathcal L)/R_y,\notag\\
 0&=AP+PA^{\mathsf T}+\mathcal Q-R_yKK^{\mathsf T},\notag\\
 K_{\ell}&=(V_z^{(\ell)} C^{\mathsf T}-\mathcal L)/R_y,\notag\\
 0&=-AV_z^{(\ell)}-V_z^{(\ell)} A^{\mathsf T}+\mathcal Q-R_yK_{\ell}K_{\ell}^{\mathsf T}.
 \label{eq:cb_care}
\end{align}
The mechanical blocks are $V=\Pi P\Pi^{\mathsf T}$ and
$V^{(\ell)}=\Pi V_z^{(\ell)}\Pi^{\mathsf T}$. The latter is the width of the
future-only likelihood.

\subsection{Spectral factor and the forward covariance}

To obtain the forward covariance from the output spectrum, we factor the transfer function from the white innovation to the readout. Let $-\mu_k$ and $-\psi_k$ be the eigenvalues of $A$ and $A-KC$,
respectively, with positive real parts for $\mu_k$ and $\psi_k$.
Let $\lambda$ denote the Laplace variable. The transfer function from the white innovation to the normalized readout is
\begin{align}
 W(\lambda)&=1+C(\lambda I-A)^{-1}K\notag\\
 &=\frac{\det(\lambda I-A+KC)}{\det(\lambda I-A)}
 =\prod_{k=1}^{N_{\rm OU}+2}\frac{\lambda+\psi_k}{\lambda+\mu_k}.
 \label{eq:cb_factor}
\end{align}
The determinant equality follows from the matrix determinant lemma.
The forward Riccati equation gives the PSD of the normalized readout,
\begin{equation}
 S_y(\nu)=2R_yW(2\pi\mathrm i\nu)W(-2\pi\mathrm i\nu)
 =2R_y|W(2\pi\mathrm i\nu)|^2.
 \label{eq:cb_psd}
\end{equation}
Here $R_y$ is the readout-increment noise intensity, $\mathbb E[dw^2]=R_y\,dt$. Its white PSD is $2R_y$. The minimum-phase factor, a stable factor with a stable causal inverse, is fixed by the complete spectrum and the normalization $W(\infty)=1$.

Define the pole differences
\begin{equation}
 e_n=\sum_k(\psi_k^n-\mu_k^n).
 \label{eq:cb_moments}
\end{equation}
In particular, $e_1$ is the difference between the first pole sums. The large-$|\lambda|$ expansion is
\begin{align}
 CK&=e_1,\notag\\
 CAK&=(e_1^2-e_2)/2,\notag\\
 CA^2K&=e_1^3/6-e_1e_2/2+e_3/3.
 \label{eq:cb_expansion}
\end{align}
These coefficients follow by exponentiating
$\log W=e_1/\lambda-e_2/(2\lambda^2)+e_3/(3\lambda^3)+
O(\lambda^{-4})$.

For the position readout $C$ in Eq.~\eqref{eq:cb_model}, the gain components satisfy
$K_q=V_{qq}/R_y$, $K_p=(V_{qp}+\varsigma_{py})/R_y$, and
$K_j=P_{qj}/R_y$, where $j$ labels an OU coordinate. The $qq$ component of the forward Riccati equation
and the second line of Eq.~\eqref{eq:cb_expansion} then give
\begin{equation}
 2\omega V_{qp}=R_yK_q^2,\qquad e_2=-2\omega \varsigma_{py}/R_y.
 \label{eq:cb_qq}
\end{equation}
To obtain the momentum variance, use
$CA^2K=\omega(-\omega K_q-\gamma K_p+\sum_jc_jK_j)$
to eliminate the auxiliary cross covariances from the $qp$ equation,
\begin{equation}
 0=\omega V_{pp}-\omega V_{qq}-\gamma V_{qp}
       +\sum_jc_jP_{qj}-R_yK_qK_p.
 \label{eq:cb_qp_care}
\end{equation}
The terms proportional to $e_1\varsigma_{py}$ cancel. The resulting mechanical
covariance is
\begin{align}
 V_{qq}&=R_y e_1,\notag\\
 V_{qp}&=\frac{R_y e_1^2}{2\omega},\notag\\
 V_{pp}&=\frac{R_y}{\omega^2}\frac{e_1^3-e_3}{3}
                  -\frac{\gamma \varsigma_{py}}{\omega}.
 \label{eq:cb_forward}
\end{align}

\subsection{Future likelihood width}

To calculate the future likelihood width, we apply the forward algebra to the reversed drift. The last two lines of Eq.~\eqref{eq:cb_care} have this form with $A\mapsto-A$ and $\mathcal L\mapsto-\mathcal L$.
The corresponding algebraic factor for the reversed drift is

\begin{equation}
 W_{\ell}(\lambda)=1+C(\lambda I+A)^{-1}K_{\ell}
             =\prod_k\frac{\lambda+\psi_k^{\ell}}{\lambda-\mu_k}.
 \label{eq:cb_future_factor}
\end{equation}
Here $-\psi_k^{\ell}$ are the stable eigenvalues of the error drift $-A-K_{\ell}C$, whereas $W_{\ell}$ has poles at $+\mu_k$.
The factorization identity
$W_{\ell}(\lambda)W_{\ell}(-\lambda)=W(\lambda)W(-\lambda)$
fixes the same squared numerator roots. Selecting the roots with positive
real part gives $\{\psi_k^{\ell}\}=\{\psi_k\}$ and hence the expansion
coefficients below.

Introduce the pole sums associated with the drift matrix $A$:
\begin{align}
 \mathcal M_1&=\sum_k\mu_k=-\operatorname{tr}A
          =\gamma+\sum_j a_j,\notag\\
 \mathcal M_3&=\sum_k\mu_k^3=-\operatorname{tr}A^3
          =\gamma^3-3\gamma\omega^2+\sum_j a_j^3.
 \label{eq:cb_open_poles}
\end{align}
The corresponding coefficients are
\[
 e_1^{\ell}=e_1+2\mathcal M_1,\qquad e_2^{\ell}=e_2,\qquad
 e_3^{\ell}=e_3+2\mathcal M_3.
\]
Relative to $W$, the denominator of $W_{\ell}$ replaces $\mu_k$ by
$-\mu_k$: odd pole sums change sign, whereas even sums do not.
Thus $\log W_{\ell}=e_1^{\ell}/\lambda-e_2/(2\lambda^2)
+e_3^{\ell}/(3\lambda^3)+O(\lambda^{-4})$.
Apply the component algebra leading to Eq.~\eqref{eq:cb_forward} with
$e_1\mapsto e_1^{\ell}$, $e_3\mapsto e_3^{\ell}$, and
$(\omega,\gamma,\varsigma_{py})\mapsto(-\omega,-\gamma,-\varsigma_{py})$.
These substitutions implement $A\mapsto-A$ and
$\mathcal L\mapsto-\mathcal L$; they give
\begin{align}
 V^{(\ell)}_{qq}&=R_y(e_1+2\mathcal M_1),\notag\\
 V^{(\ell)}_{qp}&=-\frac{R_y(e_1+2\mathcal M_1)^2}{2\omega},\notag\\
 V^{(\ell)}_{pp}&=\frac{R_y}{\omega^2}
 \frac{(e_1+2\mathcal M_1)^3-e_3-2\mathcal M_3}{3}
                  +\frac{\gamma \varsigma_{py}}{\omega}.
 \label{eq:cb_future}
\end{align}

\subsection{The conventional prescription and its scope}

To obtain the position and momentum variance biases of the conventional prescription in Eq.~\eqref{eq:conventional_rule}, subtract
Eq.~\eqref{eq:cb_forward} from Eq.~\eqref{eq:cb_future} and divide
by two:
\begin{align}
 (V_2)_{qq}-V_{qq}&=R_y\mathcal M_1,\notag\\
 (V_2)_{pp}-V_{pp}&=\frac{2\mathcal M_1}{\omega}V_{qp}
     +\frac{2\mathcal M_1^2}{\omega^2}V_{qq}\notag\\
 &\quad+\frac{R_y(4\mathcal M_1^3-\mathcal M_3)}{3\omega^2}
     +\frac{\gamma \varsigma_{py}}{\omega}.
 \label{eq:cb_bias}
\end{align}
This result retains correlated force and readout noise. Without auxiliary force variables,
$\mathcal M_1=\gamma$; with them, their decay rates contribute to the same sum.

The off-diagonal prescription requires its own calculation. Let
$\Sigma$ denote the stationary unconditional mechanical covariance.
Stationarity of $\langle q^2\rangle$, together with $\dot q=\omega p$
and no direct $q$ diffusion, implies $\Sigma_{qp}=0$. The law of total
covariance gives
\begin{equation}
 \operatorname{Cov}(f_q,f_p)=\Sigma_{qp}-V_{qp}=-V_{qp},
 \qquad (V_2)_{qp}=V_{qp}.
 \label{eq:cb_conventional_qp}
\end{equation}
Thus the conventional off-diagonal component equals $V_{qp}$ in this stationary full-band model.
\section{Bandwidth bounds and ideal thermal-noise models}
\label{app:bandwidth}
\label{sec:sm_ideal_bandwidth}

We derive the bandwidth bound for the past-conditioned center $\bm b_c$ in stationary linear-Gaussian models with matched filters and positive-definite $D=V-U$. This includes the experimental model with its effective damping. We compare the two likelihood centers and evaluate the bound in an ideal model where a thermal bath supplies both the damping and the thermal force. The model omits feedback and technical noise.

\subsection{Past-conditioned center and the omitted residual}

To identify the spectral term removed by past conditioning, we compare the two likelihood centers. Use $\bm a=\bm f-\bm s$, $D=V-U\succ0$, and $M=VD^{-1}$.
Equation~\eqref{eq:conditional_center} gives $\bm f-\bm b_c=M\bm a$.
Using Eq.~\eqref{eq:app_increment} and $M=VD^{-1}$, we obtain
$\operatorname{Cov}(\bm a,\bm f-\bm b_c)=DM^{\mathsf T}=V$.

\begingroup\displaywidowpenalty=10000
For $\bm b_0$, the unsymmetrized form
$\operatorname{Cov}(\bm f-\bm s,\bm f-\bm b_0)=V$, the mechanical projection
of Eq.~\eqref{eq:app_augmented_cross}, holds. Define
\begin{equation}
 \bm r_\perp=(\bm f-\bm b_0)-M\bm a.
 \label{eq:app_backward_residual}
\end{equation}
Then $\operatorname{Cov}(\bm a,\bm r_\perp)=0$: $M\bm a$ is the linear
least-squares projection of $\bm f-\bm b_0$ onto $\bm a$.
The residual contributes the real symmetric cross-spectrum
$\operatorname{sym}\operatorname{Re}S_{a r_\perp}$ to the three-trajectory spectral density using $\bm b_0$. Its full-band integral is zero; a restricted-band integral can be nonzero. Moreover,
\begin{equation}
 \operatorname{Cov}(\bm f-\bm b_0)
 -\operatorname{Cov}(\bm f-\bm b_c)
 =\operatorname{Cov}(\bm r_\perp)\succeq0.
 \label{eq:app_bc_projection_width}
\end{equation}
Because these two covariances are $V+V^{(\ell)}$ and $V+V_c^{(\ell)}$, Eq.~\eqref{eq:app_bc_projection_width} means that the past-conditioned likelihood is never wider than the future-only one. Either center can give the smaller band-truncation error, because the omitted cross-spectrum can either reinforce or
partly cancel the finite-band difference.\par\endgroup

\subsection{A monotone band criterion}

To bound the contribution from omitted frequencies, use the band-restricted covariance $D_B$ of $\bm f-\bm s$ defined in Eq.~\eqref{eq:conditional_band_prediction}.
Write $C_{\rm out}=D-D_B\succeq0$, and define
\begin{equation}
 \widehat D=V^{-1/2}DV^{-1/2},\qquad
 \widehat C=V^{-1/2}C_{\rm out}V^{-1/2}.
 \label{eq:app_band_whitening}
\end{equation}
Recall $g_c=\sqrt{\operatorname{cond}_2(\widehat D)}$.
Equation~\eqref{eq:conditional_band_prediction} implies
\begin{align}
 \epsilon_{B,c}&=\|\operatorname{sym}(\widehat C\widehat D^{-1})\|_2,\notag\\
 \zeta_B&=\|\widehat D^{-1/2}\widehat C\widehat D^{-1/2}\|_2.
 \label{eq:app_band_similar}
\end{align}
These transformations and those in Eq.~\eqref{eq:band_capture_metric} preserve the generalized eigenvalues of $(C_{\rm out},D)$, so they give the same $\zeta_B$.
Let $Z=\widehat D^{-1/2}\widehat C\widehat D^{-1/2}$. Since
$\widehat C\widehat D^{-1}=\widehat D^{1/2}Z\widehat D^{-1/2}$, the norm of its symmetric part is at most
$\sqrt{\operatorname{cond}_2(\widehat D)}\|Z\|_2$.
Conversely, $\widehat C\widehat D^{-1}$ has a real right eigenvector with eigenvalue
$\lambda_{\max}(Z)=\zeta_B$. The Rayleigh quotient of its symmetric
part on this vector is $\zeta_B$, which proves the lower bound in
Eq.~\eqref{eq:conditional_band_bound}.
Because $0\preceq C_{\rm out}\preceq D$, one has
$0\leq\zeta_B\leq1$. For a fixed stationary model and fixed filters, enlarging $B$ by set inclusion decreases $C_{\rm out}$ in the
positive-semidefinite order, so $\zeta_B$ is monotone nonincreasing.
The bound also covers an indefinite difference $P_{3c,B}-V$.

In matrix form, a sufficient condition for finite-band reconstruction
accuracy $\epsilon_{B,c}\leq\epsilon$ is
\begin{equation}
 D_B\succeq(1-\epsilon/g_c)D.
 \label{eq:app_sufficient_capture}
\end{equation}
The necessary condition replaces $\epsilon/g_c$ by $\epsilon$.
These matrix inequalities must hold in every direction. Small eigenvalues of
$V^{-1/2}DV^{-1/2}$ can make $g_c$ large and the sufficient bound
conservative.

\subsection{Causal thermal response and spectral conventions}
\label{sec:sm_ideal_model}

The ideal-model calculations quantify stationary band-truncation error
for structural thermal noise, quantum backaction, and independent white optical
imprecision, excluding the experiment's technical noise and fitted parameters.

The full-band calculation requires the thermal force and mechanical response to be defined over the entire frequency axis. We therefore use a causal thermal bath with finite relaxation-rate cutoffs, which specifies the response outside the range of approximately constant loss angle. With mass $m$, physical displacement $x_m$, and momentum $p_{\rm dim}=m\dot x_m$, the equations of this ideal model for $N_{\rm bath}$ relaxation elements are
\begin{align}
 m\ddot x_m&=-k_0x_m-\sum_{j=1}^{N_{\rm bath}}r_j+F_{\rm BA},\notag\\
 dr_j&=(-a_jr_j+k_j\dot x_m)dt
       +\sqrt{2k_{\rm B}T_b k_j a_j}\,dW_j.
 \label{eq:sm_maxwell_bath}
\end{align}
Here $r_j$ is the internal force of bath relaxation element $j$, $F_{\rm BA}$ is the measurement backaction force,
$k_0$ is the static spring constant, $k_j$ is that element's spring
constant, and $a_j$ is its relaxation rate. The bath temperature is $T_b$;
$k_0,k_j,a_j>0$, and the Wiener processes are independent. Without measurement backaction, the stationary joint Gibbs density is
proportional to $\exp[-E_{\rm tot}/(k_{\rm B}T_b)]$, with total energy
\begin{equation}
 E_{\rm tot}=\frac{p_{\rm dim}^2}{2m}+\frac{k_0x_m^2}{2}
      +\sum_j\frac{r_j^2}{2k_j}.
\end{equation}
This bath construction specifies the thermal noise and dissipative response together through the fluctuation--dissipation relation. Eliminating the bath coordinates gives the Laplace susceptibility and thermal force PSD
\begin{align}
 \chi(\lambda)^{-1}&=m\lambda^2+k_0+\sum_j k_j\frac{\lambda}{\lambda+a_j},\notag\\
 S_{FF,\rm th}(\nu)
 &=\sum_j\frac{4k_{\rm B}T_b k_j a_j}{a_j^2+(2\pi\nu)^2}.
 \label{eq:sm_maxwell_spectra}
\end{align}
Here $\lambda$ is the Laplace variable. The displacement variance is
$\operatorname{Var}(x_m)=\int_0^\infty S_{xx}(\nu)\,d\nu$.

With logarithmically spaced relaxation rates and
$k_j\simeq w\,\Delta\ln a_j$, where $w=2k_0\phi/\pi$, the interior of the
bath band has approximately constant loss angle $\phi$ and thermal force
PSD $4k_{\rm B}T_b k_0\phi/(2\pi\nu)$.
We use the classical thermal limit, $k_{\rm B}T_b\gg\hbar\max_j a_j$;
only the optical noises are kept at their quantum measurement limit.

The white imprecision $x_{\rm imp}$ and the white backaction force $F_{\rm BA}$ are uncorrelated, with
\begin{equation}
 S_{xx,\rm imp}S_{FF,\rm BA}
 =\frac{\hbar^2}{\eta_{\rm det}},\qquad
 S_{x_{\rm imp}F_{\rm BA}}=0,
 \label{eq:sm_optical_product}
\end{equation}
for coherent optical probing and detection efficiency $\eta_{\rm det}$ of this ideal measurement model.
The displacement-referred detector output $y_x$ has PSD
\begin{equation}
 S_{y_x}(\nu)=S_{xx,\rm imp}+
 |\chi(-2\pi\mathrm i\nu)|^2
 [S_{FF,\rm th}(\nu)+S_{FF,\rm BA}].
 \label{eq:sm_ideal_output}
\end{equation}

\subsection{Covariances and band integration}

To compare full-band and finite-band reconstruction in the ideal model, we first compute the conditional covariances. For the finite bath, continuous algebraic Riccati equations give the
forward covariance and the future-only likelihood width. The mechanical
smoother is also obtained directly from the physical spectra. We use the static-spring frequency $\omega_m=\omega_0=\sqrt{k_0/m}$ for normalization. In canonical
coordinates $q=x_m\sqrt{m\omega_m/\hbar}$ and
$p=p_{\rm dim}/\sqrt{\hbar m\omega_m}$, write $y=y_x\sqrt{m\omega_m/\hbar}$. Thus
$S_y(\nu)=(m\omega_m/\hbar)S_{y_x}(\nu)$, and the white imprecision
PSD in $q$ units is $S_{q,\rm imp}=(m\omega_m/\hbar)S_{xx,\rm imp}=2R_y$,
where $R_y$ is the readout-increment noise intensity. Define the
white-backaction momentum diffusion by
\begin{align}
 dp_{\rm BA}&=\sqrt{Q_{\rm BA}}\,dW_{\rm BA},\qquad
 \mathbb E[dW_{\rm BA}^2]=dt,\notag\\
 R_yQ_{\rm BA}&=\frac{1}{4\eta_{\rm det}}.
 \label{eq:sm_backaction_diffusion}
\end{align}
The PSD of the corresponding white momentum drive is
$2Q_{\rm BA}=S_{FF,\rm BA}/(\hbar m\omega_m)$. Let $S_q(\nu)$ be the unconditional position PSD in canonical units. The stationary smoother covariance is
\begin{align}
 U_{qq}&=\int_0^\infty
             \frac{2R_yS_q(\nu)}{S_q(\nu)+2R_y}\,d\nu,\notag\\
 U_{pp}&=\int_0^\infty
             \frac{2R_yS_q(\nu)}{S_q(\nu)+2R_y}
             \left(\frac{2\pi\nu}{\omega_m}\right)^2\,d\nu,\notag\\
 U_{qp}&=0.
 \label{eq:sm_ideal_smoother}
\end{align}
The vanishing symmetrized cross term follows from stationarity and
$p=\dot q/\omega_m$. The conditional covariance $V$ can have a nonzero position--momentum cross covariance.

\begingroup\displaywidowpenalty=10000
Let $H_f,H_s,H_{b_0}$ be the two-component transfer functions from the
position record. With $\bm a=\bm f-\bm s$, the spectrum of the difference $\bm a$ and the two band predictions are
\begin{align}
 S_{aa}(\nu)&=[H_f(\nu)-H_s(\nu)]S_y(\nu)\notag\\
 &\quad\times[H_f(\nu)-H_s(\nu)]^\dagger,\notag\\
 P_{3c,B}&=\operatorname{sym}\!\left[
       \left(\int_B\operatorname{Re}S_{aa}(\nu)d\nu\right)
       (V-U)^{-1}V\right],\notag\\
 P_{3,B}&=\int_B\operatorname{sym}\operatorname{Re}
       [(H_f-H_s)S_y(H_f-H_{b_0})^\dagger]d\nu.
 \label{eq:sm_ideal_band_integrals}
\end{align}
These expressions use the same band operator and the same physical model
for both centers. Equation~\eqref{eq:band_accuracy} measures the resulting
relative matrix error.\par\endgroup

\subsection{A fixed-bath comparison of the two centers}
\label{sec:sm_ideal_comparison}

We compare both centers in the same finite bath, with static-spring
frequency $\nu_0=\sqrt{k_0/m}/(2\pi)=\omega_0/(2\pi)$. On $[0.30,40]\nu_0$,
the largest relative error in a quadrature variance is $7.71\%$ with the
past-conditioned center $\bm b_c$ for both bath sizes. With the future-only
center $\bm b_0$, it is about $1130\%$ for $N_{\rm bath}=7$ and $468\%$
for $N_{\rm bath}=13$ ($\epsilon_B=11.3$ and $4.68$, respectively). At either bath size, only the past-conditioned center meets a $10\%$ tolerance on this band.
On $[0.05,50]\nu_0$, both bath sizes satisfy the sufficient condition in
Eq.~\eqref{eq:app_sufficient_capture}. The bound $g_c\zeta_B=0.0979$
guarantees an error below $10\%$, and the directly calculated error is
$\epsilon_{B,c}=0.0560$.

We retain the bath dispersion while keeping the static spring constant $k_0$ fixed. The
parameters are $\phi=10^{-3}$,
$n_T=k_{\rm B}T_b/(\hbar\omega_0)=10^4$, $\omega_0R_y=0.1$, and
$\eta_{\rm det}=1$, giving $Q_{\rm BA}/\omega_0=2.5$;
all damping arises from the bath. The bath has $N_{\rm bath}=7$ or $13$
logarithmically spaced rates over $a_j/\omega_0=0.01$--$100$, with
$k_j/(m\omega_0^2)=(2\phi/\pi)w_j$, where $w_j$ are logarithmic trapezoidal
weights with half weights at both endpoints. At each $N_{\rm bath}$, both
centers use the same finite bath.

\makeatletter
\def\bibsection{%
  \par\onecolumngrid@push
  \begingroup\baselineskip26\p@\bib@device{\textwidth}{245.5\p@}\endgroup
  \nobreak\@nobreaktrue\addvspace{6\p@}\par\onecolumngrid@pop
}
\makeatother

\onecolumngrid
\clearpage
\hypertarget{supplemental-material}{}
\begin{center}
\textbf{\large Supplemental Material}
\end{center}
\twocolumngrid
\makeatletter
\let\@hangfrom@section\arxivNormalHangFromSection
\let\@sectioncntformat\arxivNormalSectionCountFormat
\@removefromreset{equation}{section}
\def\theequation@prefix{}
\makeatother
\renewcommand{\theHsection}{supp.\arabic{section}}
\renewcommand{\theHsubsection}{supp.\arabic{section}.\arabic{subsection}}
\renewcommand{\theHsubsubsection}{supp.\arabic{section}.\arabic{subsection}.\arabic{subsubsection}}
\renewcommand{\theHequation}{supp.\arabic{equation}}
\renewcommand{\theHfigure}{supp.\arabic{figure}}
\renewcommand{\theHtable}{supp.\arabic{table}}

This Supplemental Material describes the calibration and noise model, the reconstruction from six intervals of one detector record, and its dependence on the analysis band. Section, equation, figure,
and table references without an S prefix refer to the main article. Bibliographic reference numbers are shared
with the main article.
The mechanical quadratures $q$ and $p$ are dimensionless position and momentum. The subscripts $qq$, $pp$, and $qp$ denote position variance, momentum variance, and symmetrized position--momentum covariance, respectively. We compare the two- and three-trajectory reconstructions with their finite-band predictions. We also compare the past-conditioned three-trajectory reconstruction with $V$, the $2\times 2$ covariance matrix of $q$ and $p$ conditioned on the measurement record through the estimation time. Complete-record simulations propagate finite-record fluctuations,
measured noise-response uncertainty, and calibration uncertainty together.
SD denotes standard deviation, SE standard error, PSD power spectral
density, ASD amplitude spectral density, RIN relative intensity noise, and OU
Ornstein--Uhlenbeck. All power and cross-spectral densities are one-sided and
expressed per hertz at positive frequency $\nu$. Transfer functions written
in angular frequency are evaluated at $\Omega=2\pi\nu$.
\setcounter{section}{0}
\setcounter{equation}{0}
\setcounter{figure}{0}
\setcounter{table}{0}
\renewcommand{\thesection}{S\arabic{section}}
\renewcommand{\thesubsection}{\thesection.\arabic{subsection}}
\renewcommand{\thesubsubsection}{\thesubsection.\alph{subsubsection}}
\renewcommand{\theequation}{S\arabic{equation}}
\renewcommand{\thefigure}{S\arabic{figure}}
\renewcommand{\thetable}{S\arabic{table}}
\makeatletter
\renewcommand{\p@subsection}{}
\renewcommand{\p@subsubsection}{}
\makeatother
\section{Experimental model, calibration, and technical noise}
\label{sec:sm_calibration}
\label{app:calibration}

The reconstruction in Sec.~\ref{sec:experiment} of the main article uses the measured detector voltage. Here we give its displacement calibration, specify how the noise sources enter the mechanical motion and readout, and compare the measured laser-frequency servo with the ideal-servo reference.

\subsection{Displacement calibration and its uncertainty}
\label{sec:sm_auxiliary_calibrations}
\label{sec:sm_calibration_dependencies}

\begin{figure*}[t]
\centering
\includegraphics[width=0.82\textwidth]{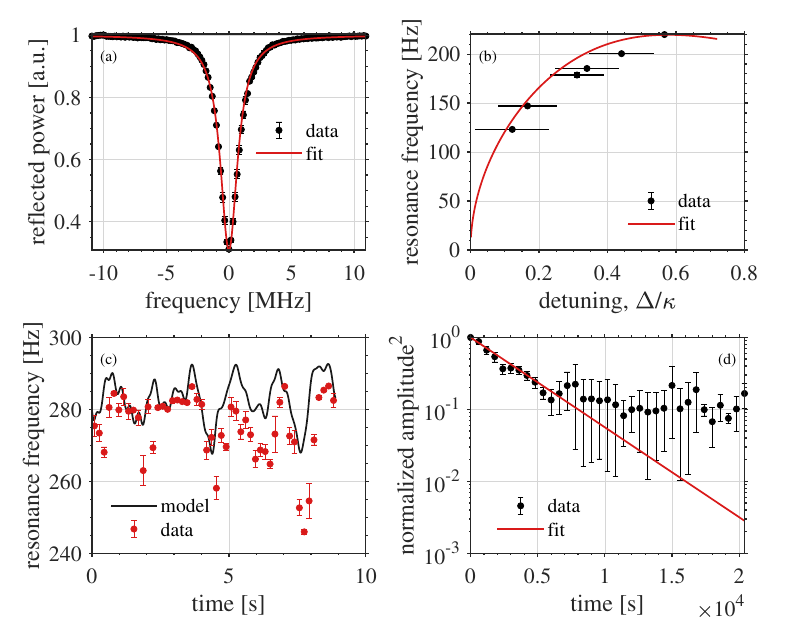}
\caption{Auxiliary calibration measurements. (a) Cavity reflection:
the points give the mean of four aligned scans and the bars span one standard error of the mean on either side.
The curve is the weighted fit including the scan mode-matching efficiency. (b) Optical-spring frequency
versus normalized detuning. Horizontal bars show the temporal detuning
standard deviation during each transfer-function measurement; vertical
bars show the resonance-fit standard error. The red curve is the
optical-spring fit with propagated detuning uncertainty. (c) Zero-crossing check on the selected
small-detuning branch. Points are means in 50 time bins with
autocorrelation-corrected standard errors; the curve is the
displacement-calibrated optical-spring prediction. (d) Suspension
ring-down. Points show the averaged normalized amplitude squared and its
standard error; the curve is the exponential fit.}
\label{fig:sm_auxiliary_calibrations}
\end{figure*}

Figure~\ref{fig:sm_auxiliary_calibrations} shows the measurements used to convert the reflected voltage to mirror displacement and set the optical parameters for each interval. The fit to the reflected-power scans includes the mode-matching efficiency of the scan. The observed resonance depth, normalized to the off-resonant reflected power, is this efficiency times the intrinsic cavity contrast $4\rho_{\rm in}(1-\rho_{\rm in})$. The fit gives the angular half width $\kappa$ and the intrinsic contrast. Here $\kappa_{\rm in}$ is the input-coupling contribution to that half width and $\rho_{\rm in}=\kappa_{\rm in}/\kappa$ its fractional contribution. The optical-spring fit in panel (b) gives the power scale $A_{\rm cal}=P_{\rm cal}\cos^2\beta$, where $P_{\rm cal}$ is the independently calibrated mode-matched input power and $\beta$ is the incidence angle. We adopt the undercoupled solution $\rho_{\rm in}=0.22246$ because it gives $\beta=(0.686\pm0.060)\,\mathrm{rad}$ ($39.3^\circ$), consistent with the design angle of $42.26^\circ$. This gives the change in cavity resonance angular frequency per unit mirror displacement, $G=2\omega_L\cos\beta/L_{\rm rt}$, where $\omega_L$ is the laser angular frequency and $L_{\rm rt}$ the cavity round-trip length.

\label{sec:sm_detuning_branch}
We use the normalized detuning $\delta=(\omega_L-\omega_{\rm cav})/\kappa$, where $\omega_{\rm cav}$ is the cavity resonance angular frequency at the operating position. At the reference mechanical frequency $283.5\,\mathrm{Hz}$, the calibrated optical spring has two candidate detunings, $\delta\simeq0.070$ and $2.1$. We estimated the oscillation frequency from successive zero crossings of the band-passed record. Its correlations with the two calibrated spring responses, compared on the same time grid after identical low-pass filtering, are $0.60$ and $-0.59$, respectively. This supports the small-detuning branch [Fig.~\ref{fig:sm_auxiliary_calibrations}(c)]. On this branch, the detuning inferred from the zero-crossing frequencies had a mean of $0.0703$ and a temporal standard deviation of $0.0034$ in units of the cavity half-linewidth. We used the mean for the reference calibration.

The suspension ring-down [Fig.~\ref{fig:sm_auxiliary_calibrations}(d)]
gives the bare mechanical resonance frequency $\nu_{\rm bare}$ and the
intrinsic quality factor $Q_{\rm bare}$. An electronic balance gave the mirror mass $m=7.71\,\mathrm{mg}$.
We used its specified resolution of $0.01\,\mathrm{mg}$ as the standard
deviation for mass in the uncertainty propagation.
For the photodiode quantum efficiency $\eta_{\rm PD}$, we used a nominal
value of $0.85$ based on the Hamamatsu G10899-03K response at
$1064\,\mathrm{nm}$ and assigned an absolute standard uncertainty of $0.08$.
The efficiency and the measured transimpedance amplifier (TIA) gain
$Z_{\rm TIA}$ determine the voltage responsivity $G_{\rm PD}$.

At the operating mechanical angular frequency $\omega_{\rm eff}$, the cavity-reflection calibration [Eq.~(5) of Ref.~\cite{PhysRevA.94.033822}] converts the measured voltage $v(t)$ to the displacement-equivalent readout $y_x(t)$ with conversion factor $\alpha$ in metres per volt,
\begin{align}
 \alpha=\frac{\sqrt{S_x(\nu)}}{\sqrt{S_v(\nu)}}
 &=\frac{\mathcal F\cos\beta}
 {\pi c m(1-\rho_{\rm in})\omega_{\rm eff}^{2}G_{\rm PD}},
 \nonumber\\
 y_x(t)&=\alpha v(t).
 \label{eq:sm_auxiliary_displacement}
\end{align}
At Fourier frequency $\nu$, $S_x$ and $S_v$ are the power spectral densities (PSDs) of $y_x$ and $v$, respectively. Here $\mathcal F$ is the cavity finesse and $c$ the speed of light. The earlier deposit labels the six processed records as displacements. Interpreting their numerical values as dimensionless quadratures gives amplitudes $8.6\%$ below those obtained from the raw voltage with the originally documented displacement calibration and zero-point length. The intermediate displacement records confirm the documented voltage-to-displacement conversion in all six intervals and locate the discrepancy in the subsequent conversion to dimensionless quadratures. For all six intervals we start from the raw detector voltage, apply Eq.~\eqref{eq:sm_auxiliary_displacement}, and then normalize the quadratures as in Eq.~\textup{(\ref{eq:exp_units})} of the main article.

Table~\ref{tab:sm_upstream} lists the calibration inputs and derived quantities. For each input condition we derive the optical coefficients and displacement calibration together and use the same values to reconstruct the observed record and to generate synthetic records. The calibration-power and displacement-detector gains are independent. We derive the intrinsic quality factor from the ring-down power-decay time $\tau_{\rm rd}$ and the bare frequency, $Q_{\rm bare}=2\pi\nu_{\rm bare}\tau_{\rm rd}$. For the fitted spring scale, the independent multiplier $r_S$ (nominal value 1, relative standard deviation $0.89\%$) accounts for the variance remaining after the contributions from $m$, $\kappa$, $L_{\rm rt}$ and $\rho_{\rm in}$ are propagated to first order. The original fit covariances were unavailable. First-order propagation gives a relative standard uncertainty of approximately $10.5\%$ for $\alpha_{\rm ref}$, the conversion evaluated at the reference mechanical frequency of $283.5\,\mathrm{Hz}$.

\begin{table*}[t]
\centering
\caption{Calibration inputs and derived quantities. For the independently sampled normal inputs, standard deviations refer to the distributions before restriction to physical domains. For $A_{\rm cal}$ and $Q_{\rm bare}$, they give the first-order uncertainty budgets used to derive their correlated values.}
\label{tab:sm_upstream}
\label{tab:sm_auxiliary_results}
\small
\begin{tabular}{lll}
\toprule
Calibration input & Nominal value & Uncertainty used in propagation \\
\midrule
Mirror mass $m$ & $7.71\,\mathrm{mg}$ & $0.01\,\mathrm{mg}$ \\
Bare frequency $\nu_{\rm bare}$ & $4.527\,\mathrm{Hz}$ & $0.002\,\mathrm{Hz}$ \\
Intrinsic quality factor $Q_{\rm bare}$ & $9.9\times10^4$ & $8.0\times10^3$ \\
Cavity half width $\kappa/(2\pi)$ & $834.8\,\mathrm{kHz}$ & $4.3\,\mathrm{kHz}$ \\
Incident power $P_{\rm in}$ & $30\,\mathrm{mW}$ & $3\,\mathrm{mW}$ \\
Photodiode efficiency $\eta_{\rm PD}$ & $0.85$ & $0.08$ \\
Round trip length $L_{\rm rt}$ & $98.81\,\mathrm{mm}$ & $0.77\,\mathrm{mm}$ \\
Input coupling $\rho_{\rm in}$ & $0.22246$ & $0.00011$ \\
Mode matching $\eta_{\rm mm}$ & $0.831$ & $0.015$ \\
Transimpedance $Z_{\rm TIA}$ & $500\,\Omega$ & $1.3\,\Omega$ \\
Spring power scale $A_{\rm cal}$ & $1.928\,\mathrm{mW}$ & $0.040\,\mathrm{mW}$ \\
Effective calibration power $P_{\rm cal}$ & $3.22\,\mathrm{mW}$ & $0.31\,\mathrm{mW}$ \\
\midrule
Derived quantity & \multicolumn{2}{l}{Calibration result and standard uncertainty} \\
\midrule
Free spectral range & \multicolumn{2}{l}{$(3.034\pm0.024)\,\mathrm{GHz}$} \\
Finesse $\mathcal F$ & \multicolumn{2}{l}{$1817\pm17$} \\
Incidence angle $\beta$ & \multicolumn{2}{l}{$(0.686\pm0.060)\,\mathrm{rad}$} \\
Cavity slope $G$ & \multicolumn{2}{l}{$(2.77\pm0.13)\times10^{16}\,\mathrm{rad\,s^{-1}\,m^{-1}}$} \\
Detector response $G_{\rm PD}$ & \multicolumn{2}{l}{$(365\pm34)\,\mathrm{V/W}$} \\
\bottomrule
\end{tabular}
\end{table*}

In verification interval $i$, we use the waveform estimate $\widehat\nu_i$ (Sec.~\ref{sec:sm_waveform_frequency}) as the local resonance frequency $\nu_i$ to determine the detuning $\delta_i$. This assumes that changes in resonance frequency arise from the optical spring at fixed feedback stiffness. Let $h(\delta)$ be the dimensionless detuning dependence of the optical spring. A subscript $0$ denotes a nominal calibration value. Each $r$ with an input as subscript is its ratio to that value, for example $r_{P_{\rm in}}=P_{\rm in}/P_{{\rm in},0}$. Then
\begin{align}
 h(\delta_i)&=\frac{\nu_i^2-\nu_{\rm bare}^2}{K_f},\qquad
 h(\delta)=\frac{\delta}{(1+\delta^2)^2},\nonumber\\
 \frac{K_f}{K_{f0}}&=r_Sr_{\eta_{\rm mm}}
                      r_{P_{\rm in}}/r_{P_{\rm cal}},\nonumber\\
 K_{f0}&=\frac{(283.5\,\mathrm{Hz})^2-\nu_{{\rm bare},0}^2}{h(0.0703)}.
 \label{eq:sm_spring_graph}
\end{align}
Here $K_f$ is the optical-spring frequency-squared scale in $\mathrm{Hz}^2$, and $K_{f0}$ is its nominal value. With $\omega_{\rm eff}=2\pi\nu_i$, the voltage-to-displacement conversion in interval $i$ is $\alpha_i=\alpha_{\rm ref}(283.5\,\mathrm{Hz}/\nu_i)^2$. The observed input remains in the fixed units $\alpha_0$ times detector voltage, where $\alpha_0$ is the nominal reference conversion. The observation matrix is divided by $\alpha_i/\alpha_0$. Before averaging intervals or comparing input conditions, we express the reconstructed and predicted covariances in common quadratures at $m_0=7.71\,\mathrm{mg}$ and $\omega_{\rm ref}=2\pi(283.5\,\mathrm{Hz})$.

\subsection{Noise sources in the mechanical motion and readout}
\label{sec:sm_optical_coefficients}

The noise budget of Fig.~\ref{fig:noise_budget_residual} of the main article separates the contributions of the modeled sources. An incident intensity fluctuation changes both the radiation-pressure force and the reflected-power readout. For a classical fractional incident-power fluctuation $u$, the force coefficient $F_u$ and the displacement-equivalent direct-readout coefficient $X_u$ are determined at the operating detuning $\delta$, together with the optical stiffness $k_{\rm opt}$,
\begin{align}
 F_u&=-\hbar G n_{\mathrm{cav}},\qquad
 X_u=\frac{R_{\rm det}}{\eta_{\rm mm}\rho_{{\rm cav},x}},\nonumber\\
 k_{\rm opt}&=\frac{2\hbar G^2n_{\mathrm{cav}}\delta}{\kappa(1+\delta^2)}.
 \label{eq:sm_optical_coefficients}
\end{align}
Here $\hbar$ is the reduced Planck constant and $n_{\mathrm{cav}}$ the mean intracavity photon number. At the reference detuning $\delta_0=0.0703$, the nominal calibration gives $n_{\mathrm{cav}}=1.1273\times10^{10}$. For the traveling wave in the ring cavity, the circulating power is the stored optical energy times the round-trip frequency, $P_{\mathrm{circ}}=\hbar\omega_L n_{\mathrm{cav}}c/L_{\rm rt}=6.385\,\mathrm W$ at $1064\,\mathrm{nm}$ and $L_{\rm rt}=98.81\,\mathrm{mm}$. The detector collects the reflected power in both the matched and unmatched spatial modes. Their combined reflectance is $R_{\rm det}=\eta_{\rm mm}\rho_{\rm cav}+1-\eta_{\rm mm}$, where $\eta_{\rm mm}$ is the mode-matching efficiency during the measurement, $\rho_{\rm cav}$ the matched-mode power reflectance and $\rho_{{\rm cav},x}$ its displacement derivative. At the small positive detuning used here, $\rho_{{\rm cav},x}<0$, so $F_u$ and $X_u$ have the same sign. The two contributions are added as amplitudes with this relative sign before the spectrum is formed.

In interval $i$, $\omega_i=2\pi\nu_i$ and $\gamma_i$ are the mechanical angular frequency and effective damping rate, and $k_{{\rm opt},i}$ is the local optical stiffness. The equations for the mirror displacement $x_m$, its physical momentum $p_{\rm dim}$ and the readout $y_x$ are
\begin{align}
 \dot x_m&=p_{\rm dim}/m,\nonumber\\
 \dot p_{\rm dim}&=-m\omega_i^2x_m-\gamma_i p_{\rm dim}+F_{\mathrm{th}}\nonumber\\
       &\quad+\sum_{a=1}^{6}f_a\xi_a^{\rm vac}+F_u u+k_{{\rm opt},i}n_{\rm fn},\nonumber\\
 y_x&=x_m+\sum_{a=1}^{6}x_a\xi_a^{\rm vac}+X_u u-n_{\rm fn},
 \nonumber\\
 n_{\rm fn}&=\frac{2\pi}{G}\delta\nu_L.
 \label{eq:sm_noise_channels}
\end{align}
Here $F_{\rm th}$ is the thermal force, $\delta\nu_L$ the laser-frequency fluctuation in hertz and $n_{\rm fn}$ its displacement equivalent. The six vacuum inputs $\xi_a^{\rm vac}$ are independent unit white noises, with force coefficients $f_a$ and direct-readout coefficients $x_a$. Four are the two quadratures at each of the input and cavity-loss ports. The fifth input describes vacuum fluctuations of the reflected unmatched mode and the sixth describes detection loss. Both have zero force coefficients. Their readout noise powers are proportional to $1-\eta_{\rm mm}$ and $(1-\eta_{\rm PD})R_{\rm det}/\eta_{\rm PD}$, respectively, with the same displacement calibration. Classical intensity noise, laser-frequency noise, thermal force and vacuum noise are mutually independent. Within each shared input, Eq.~\eqref{eq:sm_noise_channels} retains the correlation between force and readout. In particular, the same laser-frequency fluctuation $\delta\nu_L$ enters the force and readout with opposite signs.

The augmented state contains the mechanical coordinates, six thermal Ornstein--Uhlenbeck (OU) coordinates and the frequency-filter coordinates. The OU variables have exponentially decaying correlations. Their sum represents the structural thermal-force PSD $A_F/(2\pi\nu)$, whose amplitude coefficient is $A_F=4k_{\rm B}T_bm(2\pi\nu_{\rm bare})^2R_F/Q_{\rm bare}$. Here $k_{\rm B}$ is the Boltzmann constant. We use $T_b=300\,\mathrm K$ and fix $R_F=4$ across the analysis band. The observed reduction in the optically diluted quality factor was about fourfold near $280\,\mathrm{Hz}$~\cite{Matsumoto2019}. A two-mode calculation of pendulum--pitch mixing gives a factor of 4.38 for the parameters in Ref.~\cite{Sugiyama2023}. Over $130$--$2000\,\mathrm{Hz}$ the six-process representation preserves the integrated thermal-force power and differs from the target force PSD by at most $1.76\%$.

\subsection{Measured technical noise and the ideal-servo comparison}
\label{sec:sm_noise_scope}
\label{sec:frequency_noise_limits}

We use auxiliary measurements to set the incident intensity-noise level and constrain the stabilized frequency-noise spectrum for the covariance comparisons. The two intensity measurements sampled a beam picked off before the cavity, outside the intensity-stabilization loop, at times different from the mechanical record. With $R_{\rm RIN}$ the measured intensity amplitude spectral density (ASD) divided by the shot-noise ASD under the $30\,\mathrm{mW}$ input-referred convention, the one-sided classical excess PSD of the fractional intensity fluctuation $u$ is $S_u=(R_{\rm RIN}^2-1)2\hbar\omega_L/(30\,\mathrm{mW})$. Vacuum noise is included separately. The two measured ratios set the endpoints of the assumed distribution,
\begin{equation}
 R_{\rm RIN}\sim\operatorname{Uniform}(4.30,6.12).
 \label{eq:sm_rin_law}
\end{equation}
The nominal model uses the upper endpoint, $6.12$. The distribution used for uncertainty propagation has mean $5.21$.

Frequency noise drives the mirror and enters the readout directly. With $\chi_i$ the mechanical susceptibility of Eq.~\textup{(\ref{eq:exp_intensity_interference})} of the main article, the frequency-noise contributions to the PSDs of the readout $y_x$ and the mirror displacement $x_m$ are denoted by $S_{y_x,\mathrm{fn}}$ and $S_{x,\mathrm{fn}}$, respectively. Their ratio is
\begin{equation}
 \frac{S_{y_x,\mathrm{fn}}}{S_{x,\mathrm{fn}}}
 =\frac{|\chi_i k_{{\rm opt},i}-1|^2}{|\chi_i k_{{\rm opt},i}|^2}.
 \label{eq:sm_frequency_force_output}
\end{equation}
Below the optical-spring resonance the two paths partly cancel. The measured resonance sets the dynamical stiffness in $\chi_i$. The frequency-noise force coefficient $k_{{\rm opt},i}$ comes from the optical calibration in Eq.~\eqref{eq:sm_optical_coefficients}.

\begin{figure*}[t]
\centering
\includegraphics[width=0.78\textwidth]{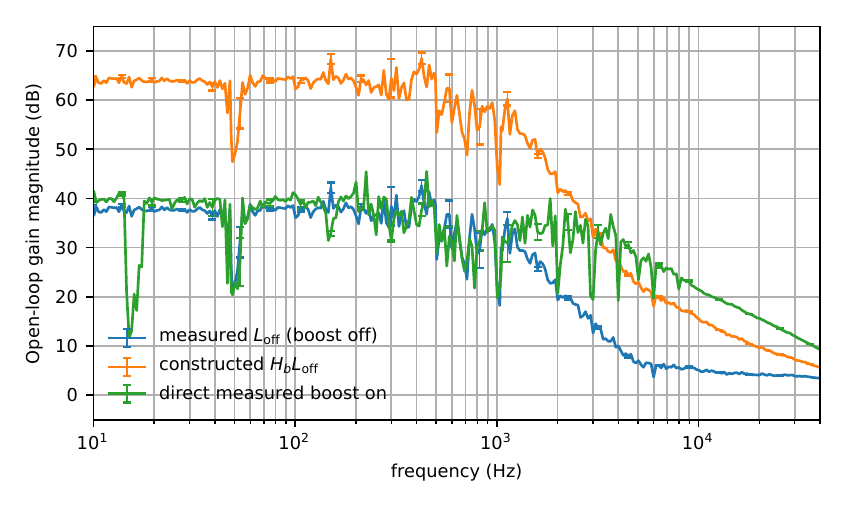}
\caption{Laser-frequency servo-gain diagnostic. The blue trace is the
measured open-loop gain $L_{\rm off}$ with the additional gain stage (boost) disabled.
The orange trace is the constructed gain $H_bL_{\rm off}$, where $H_b$ is the measured boost transfer ratio.
The green trace is the directly measured gain with boost enabled.
The plotted bars are the calibration
uncertainties associated with the gain traces. The constructed and directly measured boost-on responses differ; the adopted
model is constrained by the directly measured boost-on response.}
\label{fig:sm_servo_response}
\end{figure*}

A simultaneous out-of-loop measurement of the stabilized laser frequency was unavailable. We therefore infer its spectrum from the measured boost-on loop response (Fig.~\ref{fig:sm_servo_response}) and an assumed free-running ASD,
\begin{equation}
 \sqrt{S_{\delta\nu_L}^{\rm free}(\nu)}\simeq
 \frac{7.66\times10^3}{\nu/\mathrm{Hz}}\,\frac{\mathrm{Hz}}{\sqrt{\mathrm{Hz}}}.
 \label{eq:sm_free_frequency_asd}
\end{equation}
For each candidate spectral shape, we fixed the free-running coefficient at the geometric midpoint of the interval allowed by the measured servo response and the search bounds. We then evaluated the candidates using the observed output PSD and the five covariance comparisons in Table~\ref{tab:revised_p} of the main article. This screening informed the choice of the measured-servo model. A stable rational filter driven by white noise generates the adopted frequency noise. During uncertainty propagation we held the filter structure and free-running coefficient fixed. Across input conditions, the uncertainty in the measured servo response changes the filter poles, zeros and stabilized noise amplitude. The narrower-band covariance comparisons were reserved for a separate check.

\begin{figure*}[t]
\centering
\includegraphics[width=0.90\textwidth]{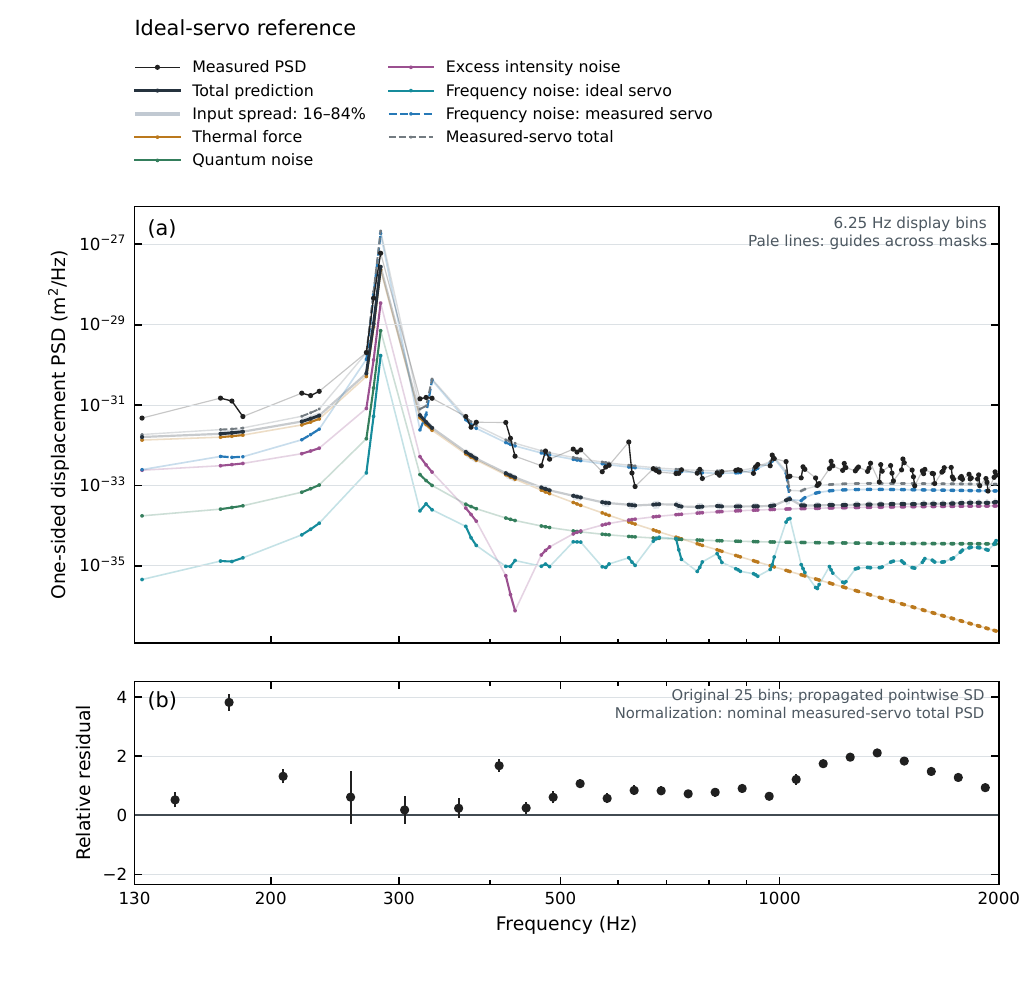}
\caption{Noise budget and residual for the ideal-servo reference
[Eq.~\eqref{eq:sm_servo_reference}], with the same data, free-running frequency
ASD, axes, and residual normalization as Fig.~\ref{fig:noise_budget_residual}.
Solid and dashed frequency-noise curves show the ideal and measured servos;
gray dashes show the measured-servo total. Shading shows pointwise 16th--84th
input percentiles of the total prediction. In (a), display bins are about
6.25 Hz. Pale lines across masks are visual guides only. In (b), the original masks and 25 bins are retained. Pointwise standard-deviation bars combine propagated input uncertainty with
finite-record fluctuations generated using the measured-servo model; the
ideal gain is fixed.}
\label{fig:sm_ideal_servo_budget}
\end{figure*}

For comparison, the ideal-servo reference uses the open-loop gain $L_{\rm off}$ measured with the boost disabled, multiplied by the measured boost transfer ratio $H_b=H_{\rm on}/H_{\rm off}$, where $H_{\rm on}$ and $H_{\rm off}$ are the boost-stage transfer functions with the boost enabled and disabled. With $L_{\rm ideal}=H_bL_{\rm off}$, the high-gain approximation gives
\begin{equation}
 S_{\delta\nu_L}^{\rm ideal}(\nu)=
 \frac{S_{\delta\nu_L}^{\rm free}(\nu)}{|L_{\rm ideal}(\nu)|^2}.
 \label{eq:sm_servo_reference}
\end{equation}
Both servo models use the same free-running ASD. In the 20 residual bins centered above $330\,\mathrm{Hz}$, the measured output PSD is $3$--$13$ times the ideal-servo prediction (Fig.~\ref{fig:sm_ideal_servo_budget}). The ideal-servo reference thus underestimates the output in this range. In the adopted model, frequency noise dominates here and its suppression is constrained by the directly measured boost-on response.

\section{Reconstruction and uncertainty of the covariance}
\label{app:implementation}

\subsection{Reconstruction from the six verification intervals}
\label{sec:sm_implementation}
\label{sec:sm_stationary_implementation}
\label{sec:sm_discrete_model}
\label{sec:sm_waveform_frequency}

The $10.01$-s record is sampled at $1$ MHz. We selected six intervals
retrospectively for comparatively stable optical-spring resonance.
The six verification intervals are 0.484--0.860, 2.200--2.360, 3.510--3.790, 5.100--5.400, 6.620--6.900, and 8.050--8.700~s.
To select a local waveform-frequency estimator, we use nineteen nonoverlapping
training windows outside the six verification intervals and their $50$-ms guards. In verification interval $i$ ($i=1,\ldots,6$), we estimate the local
resonance frequency $\widehat\nu_i$ from the phase advance of the demodulated waveform
over $4$ ms. This estimate supplies $\nu_i$ in the displacement calibration of Sec.~\ref{sec:sm_calibration_dependencies}. The local angular frequency $\omega_i=2\pi\widehat\nu_i$ sets the
effective damping rate $\gamma_i=\omega_i/Q_{\rm sh}$, where the effective
quality factor $Q_{\rm sh}=1000$ is fixed across all six intervals.

Quantum filtering estimates system observables from the measurement record up to the time of interest~\cite{Bouten2007}. In continuous quantum sensing, a time-symmetric smoother can estimate a classical signal coupled to the measured system using both earlier and later observations~\cite{Tsang2009}.

We apply notches before interval extraction and bandpass filter each
interval from $90\,\mathrm{Hz}$ to $9\,\mathrm{kHz}$. Laser frequency noise
drives the mirror and also enters the readout directly. We therefore
construct the experimental filters from the augmented model, which retains
the shared process--readout noise covariance. The stationary forward and
reverse-time transfer functions are discretized by the bilinear (Tustin) map.
Each digital filter is initialized with zero internal state at the endpoint where it starts.
In the theoretical record partition at $t_k$, the forward record includes $y_k$ and the future record starts at $y_{k+1}$. In the sampled reconstruction, both Tustin filters have a direct contribution from the processed detector sample $y_k$ at $t_k$. We apply the same filters to measured and simulated records without shifting the reverse-time output by one sample. The first and last $10\%$ of each verification interval are discarded. For the three-trajectory reconstruction
we integrate the symmetric real cross-periodogram of the trajectory
differences over $130$--$2000$ Hz. The conventional reconstruction uses
its own diagonal and cross terms [Eq.~\textup{(\ref{eq:conventional_rule})} of the main article].

\subsection{Finite-record correction and uncertainty of the comparisons}
\label{sec:sm_generating_models}
\label{sec:sm_monte_carlo}
\label{sec:sm_empirical_uncertainty}
\label{sec:sm_paired_covariance_compatibility}
\label{sec:sm_numerical_validation}

To propagate finite-record fluctuations and uncertain measured inputs,
we first draw $N_\theta=64$ input conditions. Each condition jointly
specifies the servo response, RIN, and dependent calibration quantities.
For each condition, we reconstruct the complete observed record again and
recompute the model predictions and conditional covariance in the same
fixed reference quadratures. We generate complete detector records with
the state at the end of each sampling interval and the white readout input
averaged over that interval, retaining their shared noise. We repeat local
frequency estimation, line fitting, filtering, endpoint trimming, and
spectral integration. All prescriptions process the same simulated detector record in each realization.

For each input condition $a=1,\ldots,N_\theta$, $32$ independent pilot records estimate the
mean shift caused by the finite reconstruction procedure. A separate set
of $n_e=32$ evaluation records estimates record-to-record dispersion.
The completed ensemble contains $64(32+32)=4096$ records. Let $E_j$ be the
line- and continuum-corrected reconstruction and $P_j$ its finite-band
prediction (Sec.~\ref{sec:sm_line_processing}). The conventional ($j=2$) and three-trajectory ($j=3$) prescriptions use
the future-only likelihood center $\bm b_0$. The three-trajectory estimate
with the past-conditioned center $\bm b_c$ is labeled $3c$. For each
prescription and input condition, we estimate the mean offset $\widehat\mu_{j,a}$
from the pilot records. The index $b$ numbers the records within each
simulation set. Subtracting $\widehat\mu_{j,a}$ from the experimental
reconstruction $E_{j,a}$ gives the corrected value $\widetilde E_{j,a}$.
\begin{equation}
 \widehat\mu_{j,a}=\frac{1}{32}\sum_{b=1}^{32}
   (E^{\rm pilot}_{j,ab}-P^{\rm pilot}_{j,ab}),\qquad
 \widetilde E_{j,a}=E_{j,a}-\widehat\mu_{j,a}.
 \label{eq:sm_pilot_correction}
\end{equation}
Both comparisons of $\widetilde E_{3c}$, with $P_{3c}$ and with the full-band
mechanical error covariance $V$ conditioned on the past record, use
the same correction. Because the correction is estimated from $E_{3c}-P_{3c}$ alone, the
finite-band difference $P_{3c}-V$ remains in the comparison.

For the observed record at condition $a$, the 15-component vector $\bm d_a$
collects the three entries $(qq,pp,qp)$ of each corrected residual in the five
comparisons of Table~\ref{tab:revised_p}. The corresponding vector for
evaluation record $b$ at condition $a$ is $\bm d^{\rm eval}_{ab}$. The
total uncertainty covariance $C_{\rm tot}$ for one experimental record
includes variation across input conditions and the mean record-to-record
covariance within a condition,
\begin{equation}
 C_{\rm tot}=\operatorname{Cov}_a(\bm d_a)
 +\frac{1}{N_\theta}\sum_a\operatorname{Cov}_b(\bm d^{\rm eval}_{ab}).
 \label{eq:sm_empirical_covariance}
\end{equation}
Here $\operatorname{Cov}_a$ is the sample covariance across input conditions
and $\operatorname{Cov}_b$ is the sample covariance across evaluation
records at fixed $a$. Uncertainty of the pilot mean
enters once, through the first term. To form the empirical reference
distribution, we center each contribution about its own mean,
$\overline{\bm d}$ over conditions and $\overline{\bm d}^{\rm eval}_a$
over evaluation records at each condition. Combining these centered fluctuations gives equally weighted reference
vectors $\bm e_{ab}$,
\begin{equation}
 \bm e_{ab}=
 \sqrt{\frac{N_\theta}{N_\theta-1}}(\bm d_a-\overline{\bm d})
 +\sqrt{\frac{n_e}{n_e-1}}(\bm d^{\rm eval}_{ab}-\overline{\bm d}^{\rm eval}_a)
 \label{eq:sm_total_error_reference}
\end{equation}
whose covariance is $C_{\rm tot}$. For each comparison, we use the inverse
of its three-component covariance to measure the Mahalanobis distance of the
observed mean residual from zero. A distance at or below the 95th percentile of the
corresponding reference distances places zero inside the reference region.
The model structure and selected intervals are fixed in this comparison.
Resampling the $64$ input-condition groups $5000$ times, keeping all records
within each selected group, gives the numerical stability ranges in
Table~\ref{tab:revised_p}. For the principal-variance ratio, Gaussian
purity, and ellipse radii, we propagate the full covariance of the three
reconstructed entries to first order at their mean.

\subsection{Effect of the remaining future record}
\label{sec:sm_future_information}
\label{sec:sm_future_information_diagnostics}

To assess how closely the available future record approaches the stationary
limit used in the filters, we compute the mechanical smoothing-error
covariance $U(H)$ for a remaining duration $H$. Let $P$ be the stationary
forward error covariance of the augmented state, $\Pi$ the matrix projecting
that state onto the two mechanical coordinates, and $J(H)$ the information
matrix for the augmented state supplied by a future record of duration $H$
(Appendix~\ref{app:identity}). With $\mathsf T$ denoting transpose,
\begin{equation}
 U(H)=\Pi[P^{-1}+J(H)]^{-1}\Pi^{\mathsf T},\qquad J(0)=0.
 \label{eq:finite_horizon_covariances}
\end{equation}
Additional future observations reduce or preserve the error variance in
every direction as $U(H)$ approaches its stationary value $U$.

The check covers the $64$ input conditions and six intervals ($384$ cases) of the adopted model. At the latest
retained times, $16$--$65$ ms of future record remain. Even there, the
future record reduces the forward error variance by at least $18.8\%$ in
every mechanical direction. The relative deviation $\|U^{-1/2}[U(H)-U]U^{-1/2}\|_2$, measured in the
spectral norm, is at most $16.9\%$ at the latest retained times and $0.333\%$ at the interval midpoints. A separate check of the stationary future-only
likelihood finds that its variance exceeds the forward variance by a
factor of $10^7$ in the least informative direction of each case. We
retain those directions when forming the future-only center. The
stationary joint moments satisfy the full-band three-trajectory identity
to numerical precision for both centers.

\section{Output spectra and bandwidth dependence of the reconstruction}
\label{sec:sm_spectra_bandwidth}

\subsection{Line processing and spectral weights}
\label{sec:sm_line_processing}
\label{sec:sm_diagnostics}
\label{sec:sm_subbands}

Figures~\ref{fig:interval_psd} and \ref{fig:noise_budget_residual} compare the measured and predicted spectra across frequency. The covariance comparisons in Table~\ref{tab:revised_p} use the same record but weight each frequency according to the trajectory filters.

Removing mains harmonics also attenuates the broadband signal used to reconstruct the covariance. The index $j=2,3,3c$ labels the conventional, three-trajectory, and past-conditioned three-trajectory reconstructions, respectively (Sec.~\ref{sec:sm_monte_carlo}). Each integral over $B$, the range of positive Fourier frequencies included in the analysis, is a $2\times2$ matrix in dimensionless position $q$ and momentum $p$. Let $E_{j,\rm proc}$ be the integral from the processed record and $P_{j,\rm proc}$ the predicted broadband continuum integral with the same notch and bandpass responses and Hann windows. Let $P_{j,\rm cont}$ be the unprocessed continuum prediction over $B$. We fit the harmonic waveform to the complete raw record and compute its contribution $\widehat L_j$ with the same processing and integration as for the data. The continuum-corrected statistic and its prediction are
\begin{align}
 E_j&=E_{j,\rm proc}+P_{j,\rm cont}-P_{j,\rm proc}-\widehat L_j,\nonumber\\
 P_j&=P_{j,\rm cont}.
 \label{eq:sm_continuum_correction}
\end{align}
The unprocessed prediction $P_{j,\rm cont}$ cancels in the residual $E_j-P_j=E_{j,\rm proc}-P_{j,\rm proc}-\widehat L_j$. Each complete-record simulation refits the harmonics, including the continuum power that the fit can absorb. The periodograms in Fig.~\ref{fig:interval_psd} omit bins within $\max(2/T_i,5\,\mathrm{Hz})$ of each fitted harmonic, where $T_i$ is the full duration of interval $i$. The covariance integrals use Eq.~\eqref{eq:sm_continuum_correction} before display rebinning.

With the stationary-model filters held fixed, let $S_y(\nu)$ be the scalar one-sided PSD per hertz of the normalized detector record $y$ at Fourier frequency $\nu$. Here $y$ is the calibrated displacement readout $y_x$ in units of $\sqrt{\hbar/(m\omega_i)}$ [main-text Eq.~\textup{(\ref{eq:exp_units})}]. The two-component column transfer functions $H_f$, $H_{b_0}$, and $H_s$ map this record to the forward trajectory $\bm f$, future-only likelihood center $\bm b_0$, and smoothed trajectory $\bm s$, respectively. The real part of the three-trajectory cross-spectrum matrix $S_3(\nu)$ [main-text Eq.~\textup{(\ref{eq:three_spectrum})}] is the output PSD multiplied by the weight matrix $W_3(\nu)$. For the trajectory differences $\bm f-\bm s$ and $\bm f-\bm b_0$, the transfer functions are $H_a=H_f-H_s$ and $H_d=H_f-H_{b_0}$. Taking $\operatorname{Re}$ entry by entry and writing $\dagger$ for conjugate transpose gives
\begin{align}
 W_3(\nu)&=\tfrac12\operatorname{Re}\!\left[H_aH_d^\dagger+H_dH_a^\dagger\right],\notag\\
 \operatorname{Re}S_3(\nu)&=W_3(\nu)S_y(\nu).
 \label{eq:main_spectral_weights}
\end{align}
Integrating $W_3 S_y$ over $B$ predicts the three-trajectory reconstruction in that band. The $qq$, $pp$, and $qp$ entries correspond to position variance, momentum variance, and their cross covariance. The diagonal entries of $W_3$ can have either sign. The conventional diagonal weights $W_{2,qq}=|H_{d,q}|^2/2$ and $W_{2,pp}=|H_{d,p}|^2/2$ are nonnegative. For the past-conditioned center $\bm b_c$, its transfer function replaces $H_{b_0}$. A local discrepancy in the output PSD can have little effect on an integrated covariance entry if its spectral weight is small or if weighted residuals cancel across frequencies.

\subsection{Bandwidth dependence of the experimental reconstruction}
\label{sec:sm_bandwidth_scan}

Figure~\ref{fig:bandwidth_scan} compares the reconstructed $qq$, $pp$, and $qp$ entries with the corresponding entries of two matrices. For the past-conditioned three-trajectory reconstruction, $P_{3c,B}$ is the model prediction integrated over band $B$. The full-band mechanical covariance conditioned on the past record, $V$, is common to all five bands. The lower edge is fixed at $130\,\mathrm{Hz}$ and the upper edge varies over $200$, $300$, $500$, $1000$, and $2000\,\mathrm{Hz}$. The six intervals, 64 input conditions, reference quadratures, filters, and preprocessing are common to all five bands.

\begin{table}[tb]
\caption{Measured change minus predicted change when the upper band edge is extended. Each change is the difference between two adjacent cumulative bands in Fig.~\ref{fig:bandwidth_scan}. Standard uncertainties include the correlations between those bands. The final column is the largest absolute residual divided by its standard uncertainty among the three entries.}
\label{tab:sm_bandwidth_contrasts}
\centering\small\setlength{\tabcolsep}{1.5pt}
\begin{tabular*}{\columnwidth}{@{\extracolsep{\fill}}lrrrr@{}}
\toprule
\shortstack{Added slice\\($\mathrm{Hz}$)} & $qq$ & $pp$ & $qp$ & \shortstack{Max.\\ratio} \\\midrule
$200$--$300$ & $37.9\pm50.2$ & $90.3\pm74.1$ & $48.6\pm43.2$ & $1.22$ \\
$300$--$500$ & $-44.8\pm82.9$ & $-3.9\pm67.5$ & $-19.9\pm61.6$ & $0.54$ \\
$500$--$1000$ & $6.6\pm49.5$ & $6.2\pm29.4$ & $5.5\pm33.6$ & $0.21$ \\
$1000$--$2000$ & $89.4\pm44.4$ & $31.0\pm15.7$ & $48.1\pm23.8$ & $2.03$ \\
\bottomrule
\end{tabular*}
\end{table}

For each input condition and band, Gaussian second moments give the expected finite-record integral. We subtract this expectation and the fitted-line contribution from the observed integral, then add $P_{3c,B}$. At $130$--$2000\,\mathrm{Hz}$ this analytic correction and the Monte Carlo pilot correction used in Fig.~\ref{fig:revised_covariance} and Table~\ref{tab:revised_covariance} give reconstructions that differ by at most $0.14$ times the corresponding component standard uncertainty.

The five bands share the same data. For the three entries in each of the five bands, we form a $15\times15$ uncertainty covariance matrix by adding the sample covariance across the 64 input conditions to the average finite-record covariance obtained from Gaussian fourth moments. Band increments use this joint covariance. Differences from $P_{3c,B}$ and $V$ retain the shared input dependence of reconstruction and prediction. The scan uses a stationary notch response, holds the fitted frequencies and lines fixed, and treats the six intervals as independent. At $130$--$2000\,\mathrm{Hz}$ the total component standard uncertainties agree with the complete-record Monte Carlo values within $1.8\%$, supporting the componentwise error bars at this bandwidth. The total variance along correlated component directions differs by up to $22\%$, so for the joint three-component comparisons in Table~\ref{tab:revised_p} we use the empirical reference distributions from the complete-record simulations. For the ellipse area in Fig.~\ref{fig:bandwidth_scan}(d), the mean matrices use all input conditions. First-order propagation gives area standard uncertainties of $27$--$36\%$ of the mean areas. For the measured and predicted area changes, we propagate the correlations between bands and their covariance due to shared input conditions.

In the $130$--$2000\,\mathrm{Hz}$ band, all three entries are within $0.4$ standard uncertainties of $V$. Table~\ref{tab:sm_bandwidth_contrasts} shows what changes when each frequency range is added. Through $1000\,\mathrm{Hz}$ the measured and predicted increments agree within $1.3$ standard uncertainties in every entry. Adding $1000$--$2000\,\mathrm{Hz}$ gives an excess of about two standard uncertainties in each entry, with correlations of $0.96$--$0.99$ among the three residuals.

\renewcommand{\bibsection}{\par\addvspace{6pt}}

\end{document}